\documentclass[fleqn,usenatbib]{mnras}

\usepackage{newtxtext,newtxmath}

\usepackage[T1]{fontenc}
\usepackage{soul}

\DeclareRobustCommand{\VAN}[3]{#2}
\let\VANthebibliography\thebibliography
\def\thebibliography{\DeclareRobustCommand{\VAN}[3]{##3}\VANthebibliography}

\usepackage{graphicx}	
\usepackage{amsmath}	
\usepackage{academicons}
\newcommand*{\Grad}{\vec{\nabla}}

\newcommand*{\Curl}{\vec{\nabla}\!\times\!}
\renewcommand*{\vec}[1]{\ensuremath{\boldsymbol{#1}}}

\title[Magneto-thermal evolution in the cores of adolescent neutron stars]{Magneto-thermal evolution in the cores of adolescent neutron stars:\\ The Grad-Shafranov equilibrium is never reached in the `strong-coupling' regime}
\author[N. Moraga et al.]{
Nicol\'as A. Moraga,$^{1}$\thanks{E-mail: nicolas.moraga@ug.uchile.cl}
Francisco Castillo,$^{2}$
Andreas Reisenegger,$^{2}$
Juan A. Valdivia,$^{1}$
and Mikhail E. Gusakov$^{3}$
\\
$^{1}$ Departamento de F\'isica, Facultad de Ciencias, Universidad de Chile, Casilla 653, 7800024, Santiago, Chile\\
$^{2}$ Departamento de F\'isica, Facultad de Ciencias B\'asicas, Universidad Metropolitana de Ciencias de la Educaci\'on, Av. Jos\'e Pedro Alessandri 774,\\
Nuñoa, Santiago, Chile\\
$^{3}$ Ioffe Institute, Polytekhnicheskaya 26, 194021 St Petersburg, Russia
}

\date{Accepted XXX. Received YYY; in original form ZZZ}

\pubyear{2023}

\begin{document}
\label{firstpage}
\pagerange{\pageref{firstpage}--\pageref{lastpage}}
\maketitle

\begin{abstract}
At the high temperatures present inside recently formed neutron stars ($T\gtrsim 5\times 10^{8}\, \text{K}$), the particles in their cores are in the "strong-coupling" regime, in which collisional forces make them behave as a single, stably stratified, and thus non-barotropic fluid. In this regime, axially symmetric hydromagnetic quasi-equilibrium states are possible, which are only constrained to have a vanishing azimuthal Lorentz force. In such equilibria, the particle species are not in chemical ($\beta$) equilibrium, so $\beta$ decays (Urca reactions) tend to restore the chemical equilibrium, inducing fluid motions that change the magnetic field configuration. If the stars remained hot for a sufficiently long time, this evolution would eventually lead to a chemical equilibrium state, in which the fluid is barotropic and the magnetic field, if axially-symmetric, satisfies the non-linear Grad-Shafranov equation. In this work, we present a numerical scheme that decouples the magnetic and thermal evolution, enabling to efficiently perform, for the first time, long-term magneto-thermal simulations in this regime for different magnetic field strengths and geometries. Our results demonstrate that, even for magnetar-strength fields $\gtrsim 10^{16} \, \mathrm{G}$, the feedback from the magnetic evolution on the thermal evolution is negligible. Thus, as the core passively cools, the Urca reactions quickly become inefficient at restoring chemical equilibrium, so the magnetic field evolves very little, and the Grad-Shafranov equilibrium is not attained in this regime. Therefore, any substantial evolution of the core magnetic field must occur later, in the cooler "weak-coupling" regime ($T\lesssim 5\times 10^8 \, \mathrm{K}$), in which Urca reactions are effectively frozen and ambipolar diffusion becomes relevant.

\end{abstract}




\section{Introduction}

Neutron stars (NSs) are grouped into several classes according to their very different surface magnetic field strengths, rotational periods, and the presence or absence of accretion from a binary companion. These properties vary noticeably; for example, millisecond pulsars (MSPs) have surface magnetic fields that typically range around $\sim 10^{8-9}\,\text{G}$, whilst in magnetars these values can reach $\sim 10^{14-15}\,\text{G}$ at their surface \citep{kaspi2017magnetars}, with possibly even stronger internal fields. Therefore, achieving a detailed comprehension of the different processes that lead to the evolution of the magnetic field of NSs is a key step to understand their observed phenomenology and how some of the different NS classes could be related to each other.

The problem of how the NS magnetic field evolves has been deeply studied in the literature throughout the last decades, from the seminal paper by \cite{goldreich1992magnetic} to a substantial amount of recent work that approached the problem either theoretically or numerically, given its complicated and non-linear characteristics (see 
\citealt{Pons_2019} for a recent review). Among all the mechanisms that promote the magnetic evolution, by far the best understood are those occurring in the NS crust, where ions have very restricted mobility (unless the magnetic field is strong enough to break the crust), so the evolution depends only on the motion of 
the electrons. Therefore, the long-term mechanisms that control the evolution of the field in this region are Ohmic diffusion, i.~e., current dissipation by electric resistivity; and Hall drift, which corresponds to advection of the magnetic field lines by the electron fluid motion. Though Hall drift by itself does not dissipate magnetic energy, it changes the magnetic field configuration, creating small-scale structures that dissipate more quickly, and, together with Ohmic dissipation, it may lead to a special, quasi-steady magnetic field configuration called a ``Hall attractor'' 
\citep{Cumming_2004,Gourgoliatos2014}.

In the NS core, the physics is more complex and not yet well understood. The NS core is a mixture of neutrons (mostly), protons, and electrons, joined by muons and other, more exotic species at increasing densities. Also, it is generally accepted that neutrons and protons can become superfluid and superconducting respectively, at temperatures $T\sim 10^{8}-10^{10}\,\text{K}$ \citep{migdal1959superfluidity,page2013stellar}, leading to a complex dynamics \citep{glampedakis2011magnetohydrodynamics, gd16, Gusakovetal2020,DommesGusakov2021}. 
In the present paper, we will restrict ourselves to the case of a ``normal'' core, i.~e., non-superconducting and non-superfluid, which is likely to be realistic for the high temperatures and strong magnetic fields that our work will mostly focus on.

After the supernova explosion, a proto-NS is born in an extremely hot, liquid state  
opaque to neutrinos. However, within a minute, the star becomes neutrino-transparent and the young NS is formed \citep{burrows1986,Keil1995,Pons_1999,lander221protoNS}. 
At nearly the same time, the magnetic field reaches an equilibrium configuration that is likely to fill the entire volume of the NS, as in the simulations of \citet{braithwaite2004fossil} and \citet{Becerra22a}. This configuration is stabilized by the composition gradient of matter (radially decreasing proton-to-neutron ratio) within the NS core, which causes radial buoyancy forces that oppose convective motions
\citep{pethick1992,Reisenegger92,goldreich1992magnetic}. On the other hand, since the electric conductivity of the NS core is very high, the magnetic field is effectively frozen into the charged particles. Thus, any process that promotes the magnetic evolution has to move the charged particles in such a way that the stable stratification is overcome. This can be achieved in different ways in the so-called ``strong-coupling'' and ``weak-coupling'' regimes, which are effective in different temperature ranges \citep{goldreich1992magnetic,hoyos2008magnetic,reisenegger2009stable,gusakov2017evolution,castillo2020twofluid}:
\begin{enumerate}
    \item At high temperatures, all particle species in the core are strongly coupled to each other due to frequent 
    collisions. As a result, the charged particles and neutrons move together as a single fluid that is strongly constrained by buoyancy forces as a consequence of stable stratification. However, by the effect of $\beta$-decays (so-called ``Urca-reactions'', e.~g., \citealt{shapiro1983physics}), the fluid elements can gradually adjust their chemical composition, so that the stable stratification is overcome, and the fluid can transport the magnetic flux. 
    Therefore, the dynamics of this ``strong-coupling'' regime is characterized by bulk motions where matter behaves as a single, stably stratified, non-barotropic fluid (i.~e., the pressure depends on the non-uniform chemical composition in addition to density), where the interplay between the magnetic field dynamics and Urca reactions determines the time scale over which the magnetic field evolves. 
    
    \item As the star cools, the NS core transits into the ``weak-coupling'' regime, in which the collisional coupling has decreased enough to make ambipolar diffusion (the relative motion of charged particles with respect to neutrons) possible, while the rates of the Urca reactions quickly drop, preventing further conversions between the particle species. In this regime, the finite relative velocity of neutrons and charged particles, although smaller than the bulk flow velocity \citep{ofegeim2018,castillo2020twofluid}, is crucial to allow the magnetic field to evolve and determines the rate at which magnetic energy is dissipated through collisions between neutrons and charged particles. 
    This process heats the NS core and has been invoked to explain the activity of magnetars due to its strong dependence on the magnetic field intensity \citep{DuncanThompson1995,DuncanThompson1996,beloborodov2016,Tsuruta2023}. 
    It has also been suggested to explain the low magnetic fields of MSPs, as the collisional coupling  decreases at low temperatures, so the time scale of ambipolar diffusion may become short enough to cause substantial magnetic field decay in old NSs before their spin-up by accretion \citep{cruces2019}. However, it is important to note that this scenario ignores the effects of superfluidity and superconductivity, which may have a significant impact on the dissipation of magnetic energy in the core of old NSs
    (see e.~g. \citealt{KantorGusakov2018,Gusakovetal2020}).
\end{enumerate}    

We note that this kind of dynamics (especially for the strong-coupling regime) was not correctly described by \citet{goldreich1992magnetic} and \citet{DuncanThompson1996}, who treated the neutrons as a fixed background and thus required the motion of the charged particles to overcome the friction force due to proton-neutron collisions, which is very strong at high temperatures, but which does not act if all particles move together. A similar approach was also adopted by \citet{Passamonti2017ambipolar,vigano2021} and later by \citet{igoshev2023}. 

The formal model with moving neutrons has been discussed by several authors, for instance, \citet{su95,Reisenegger2005,reisenegger2009stable}. In these studies, the velocity of the neutrons was not explicitly calculated. The original method for calculating all velocities of different particle species for a given stellar magnetic field 
 configuration was proposed in \citet{gusakov2017evolution}. That work demonstrated 
 that a magnetic field inevitably leads to the emergence of large-scale flows of stellar matter, 
 which influence the evolution of the magnetic field.
 Further, in the study by \citet{ofegeim2018}, the velocity of these flows was explicitly calculated for a range of magnetic field models. It was shown that the velocity of matter as a whole can 
 significantly exceed the relative (diffusion) particle velocities, which were calculated in 
 previous studies. A conclusion was drawn about the substantial acceleration of magnetic 
 evolution when accounting for this (previously overlooked) effect.
 These findings were later confirmed 
 in the detailed self-consistent numerical simulations of the magnetic field evolution by \citet{castillo2020twofluid} (see below for details).

\cite{Castillo2017} performed the first simulations that evolved the magnetic field and the small density perturbations it induces on the charged particles inside a NS core in axial symmetry (2 dimensions). This model did not include Urca reactions and assumed constant friction coefficients (thus implicitly constant temperature) and motionless neutrons, which represents a simplified version of the weak-coupling regime. Later, \citet{castillo2020twofluid} included the motion of the neutrons as well as the different density profiles of neutrons and charged particles (with a toy-model equation of state), thus explicitly accounting for the stable stratification, while still focusing on the weak-coupling regime at constant temperature.

Despite the improvements that have gradually been implemented in these papers, an important limitation persists within their numerical framework as the quotient between the time scales of interest is proportional to the ratio of magnetic to fluid pressure, which is $\lesssim 10^{-6}$ for realistic magnetic fields. Thus, unrealistically large magnetic field strengths are needed to run the simulations for a long enough time. In addition, the crucial influence of the thermal evolution on the dynamics of the NS core has not been considered in the contexts of the strong-coupling regime, or in the weak-coupling regime with independent velocity fields for neutrons and charged particles. This work can be considered as the continuation of the previous papers, addressing these issue. 

The aim of this work is to report the first long-term simulations of the magneto-thermal evolution of a NS core in the strong-coupling regime. We present a new numerical scheme where, by neglecting the time derivatives in the continuity equations, the quotient between the time scales of interest becomes independent of the magnetic field strength, allowing to perform simulations with realistic parameters and thus speeding up the simulation code considerably. Also, we apply a strategy that allows us to separate the magnetic and thermal evolution in the equations, in which we first run the evolution of the magnetic field at constant temperature and then introduce the thermal evolution through a change in the time variable  
(see \S~\ref{sec:Tev} for more details).

This paper is structured as follows: In \S~\ref{secModel}, we present the main equations and assumptions governing the physical model, which is to be numerically solved in axial symmetry. We discuss the time scales associated with the problem and describe the application of the novel numerical scheme to the strong-coupling regime. In \S~\ref{results_contantT}, we present the outcomes of our simulations initially under constant temperature conditions. Subsequently, in \S~\ref{sec:Tev}, we
describe the strategy to include the thermal evolution and present the results of the magneto-thermal evolution. Lastly, in \S~\ref{sec:conclusion},
we summarize our results and outline the main conclusions.

\begin{figure*}
    \includegraphics[width=11cm]{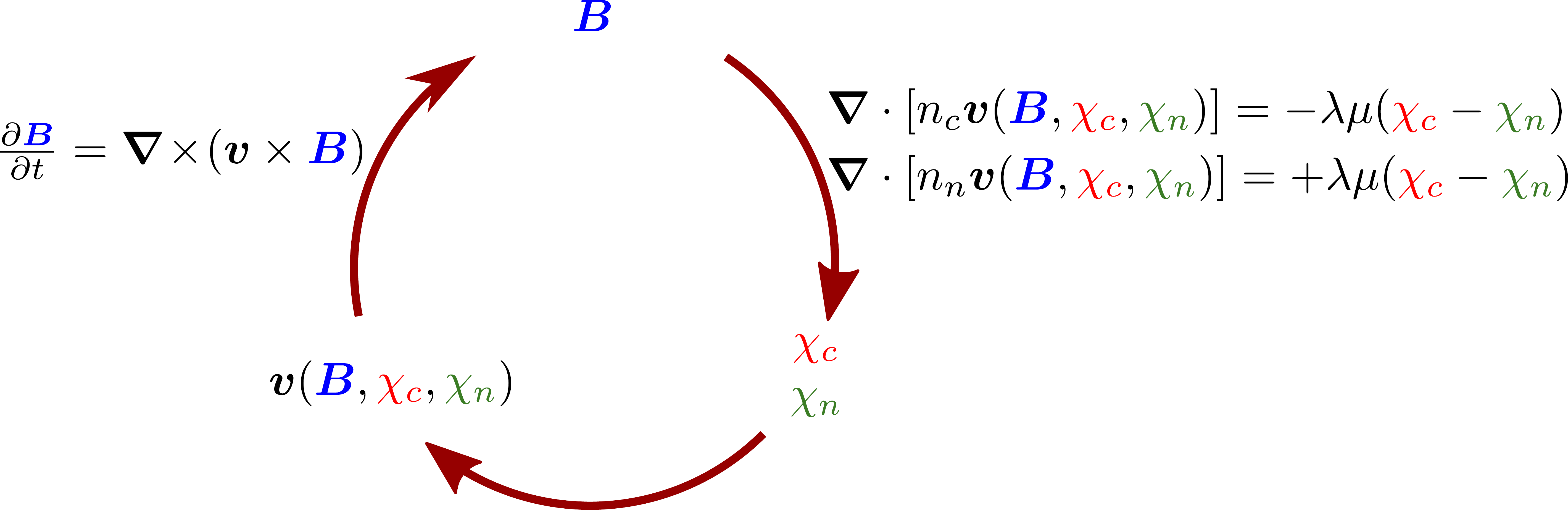}
    \caption{Schematic representation of the self-consistent method implemented to evolve the magnetic field in the strong-coupling regime.}
    \label{fig1}
\end{figure*}

\section{Model for the evolution of the magnetic field}\label{secModel}

\subsection{General equations}\label{secGeneralEquations}

Following the model of \citet{castillo2020twofluid}, 
we consider a spherically symmetric, non-rotating, non-magnetic background NS core composed of neutrons ($n$), protons ($p$), and electrons ($e$) in chemical and hydrostatic equilibrium, i.~e.
\begin{gather}   
\mu_{n}=\mu_{c}\equiv\mu(r),\label{eqChemEq}\\
  \Grad \mu+\dfrac{\mu}{c^{2}}\Grad \psi= 0,\label{eqHDEqui}
\end{gather}
where the radial functions $\psi(r)$ and $\mu_{i}(r)$ ($i=n,c$) are the gravitational potential and the chemical potentials, respectively. The subscript $c$ denotes charged particles, i.~e., $\mu_c\equiv\mu_e+\mu_p$.  

The presence of the magnetic field perturbs this equilibrium. However, since the ratio between the magnetic pressure and degeneracy pressure $P$ in NS interiors is $B^{2}/8\pi P \lesssim 10^{-6}$ \citep{reisenegger2009stable}, the perturbations induced by the magnetic field are expected to be of the same order. Thus, the chemical potentials and particle densities can be split into a time-independent, spherically symmetric background and a much smaller, time-dependent perturbation:
\begin{align}
	 n_i(\vec{r},t) & = n_i(r) + \delta n_i(\vec{r},t),\quad  (i=n,c), \\
	 \mu_i(\vec{r},t) & =\mu(r) + \delta \mu_i(\vec{r},t),
\end{align}
where the condition of charge neutrality allows to write $n_p(\vec r, t)=n_e(\vec r, t)\equiv n_c(\vec r, t)$. 
The number density perturbations are related to the chemical potential perturbations by 
\begin{equation}\label{eqPerturb}
    \delta \mu_{i}= \sum_{j=n,c} K_{ij}\delta n_{j}; \quad K_{ij}=K_{ji}=\dfrac{\partial \mu_{i}}{\partial n_{j}},
\end{equation}
and satisfy baryon number conservation: 
\begin{equation}\label{eqBaryon}
    \int_{\cal{V}} (\delta n_n+\delta n_c)d^{3}x=0,
\end{equation}
where $\cal{V}$ is the core volume.
For simplicity, we consider an axially symmetric magnetic field, which can be decomposed as
\begin{equation}\label{eqPoloidaToroidalB}
    \vec{B} = \Grad \alpha \times \Grad \phi+\beta \Grad\phi,
\end{equation}
where the scalar potentials $\alpha(t,r,\theta)$ and $\beta(t,r,\theta)$ generate the poloidal (meridional) and toroidal (azimuthal) magnetic field components, respectively.
Here, $r$ is the radial coordinate, and $\theta$ and $\phi$ are the polar and azimuthal angles, respectively, so $\Grad\phi=\hat{\phi}/(r\sin \theta)$, where $\hat\phi$ is the unit vector in the $\phi$ direction. The functions $\alpha(r,\theta,t)$ and $\beta(r,\theta,t)$ are 
known as the poloidal flux and poloidal current functions, respectively, because
$2\pi \alpha(r,\theta,t)$ is the magnetic flux and $c\beta(r, \theta, t)/2$ is the electric current enclosed by an azimuthal circle at given $r$ and $\theta$. 
The curves $\alpha=\text{const}$ are the poloidal magnetic field lines.

In order to study the magnetic field evolution, we need to integrate 
the Faraday induction equation 
\begin{equation}\label{eqFaraday}
    \dfrac{\partial \vec{B}}{\partial t}=\Grad\times\left(\vec{v}_{c}\times \vec{B}\right),
\end{equation}
where $\vec{v}_{c}=\vec{v}_{n}+\vec{v}_{ad}$ is the charged-particle velocity, obtained as the sum of the neutron velocity $\vec{v}_n$ and the ambipolar velocity $\vec{v}_{ad}$ (relative motion of charged particles with respect to neutrons). Using the poloidal-toroidal decomposition (equation \ref{eqPoloidaToroidalB}),
equation (\ref{eqFaraday}) can be written as the following coupled equations for $\alpha$ and $\beta$:
\begin{gather}
    \dfrac{\partial \alpha}{\partial t}= -\vec{v}_{c}\cdot \Grad \alpha,\label{eqFaradayA}\\
     \dfrac{\partial \beta}{\partial t}=r^{2}\sin^{2} \theta \, \Grad \cdot \left[ \dfrac{\left(\vec{v}_{c}\times \vec{B}\right)\times \hat{\phi}}{r \sin \theta}\right]\label{eqFaradayB}.
\end{gather}
Therefore, in order to evolve the magnetic field, we need to compute $\vec{v}_{c}$, which is obtained from the sum of 
\begin{equation}
        \vec{v}_{n}= \dfrac{1}{\zeta n_{n}}\left(\vec{f}_{B}+\vec{f}_{n}+\vec{f}_{c}\right),\label{eqVn}
\end{equation}
and
\begin{equation}
     \vec{v}_{ad}=\dfrac{1}{\gamma_{cn}n_{c}n_{n}}\left(\vec{f}_{B}+\vec{f}_{c}\right).\label{eqVad}
\end{equation}
The first term in the parenthesis on the right-hand side of equation (\ref{eqVn}) is the Lorentz force, 
\begin{equation}\label{eqfB}
    \vec{f}_{B}=\frac{(\Curl \vec{B})\times \vec{B}}{4\pi},
\end{equation}
whose poloidal and toroidal components can be written as
\begin{equation}\label{eqfBPol}
\vec{f}_{B}^{\text{Pol}}=-\frac{\Delta^*\alpha\Grad\alpha+\beta\Grad\beta}{4\pi r^2\sin^2\theta}
\end{equation}
and 
\begin{equation}\label{eqfBTor}
    \vec{f}_{B}^{\text{Tor}}=\frac{\Grad\beta\times\Grad\alpha}{4\pi r^2\sin^2\theta},
\end{equation}
respectively, where  
\begin{align}
    \Delta^{*} & \equiv r^{2}\sin^{2}\theta\, \Grad \cdot\left(\dfrac{1}{r^{2}\sin^{2}\theta  }\Grad\right) \\\nonumber & = \dfrac{\partial^{2} }{\partial r^{2}}+\dfrac{\sin \theta}{r^{2}}\dfrac{\partial}{\partial \theta}\left(\dfrac{1}{\sin \theta}\dfrac{\partial}{\partial \theta}\right)
\end{align}
is the ``Grad-Shafranov (GS) operator''. The other two terms include the forces due to degeneracy pressure and gravity on each of the particle species (neutrons and charged particles) and can be written as 
\begin{equation}
    \vec{f}_i=-n_i \mu \Grad\chi_i,
\end{equation} 
with 
$\chi_i\equiv\delta \mu_i/\mu$ ($i=n,c$). The latter forces are purely poloidal since we are assuming axial symmetry. Throughout this work, we use the Cowling approximation, i.~e., we do not consider the perturbations of the gravitational potential.

The parameter $\zeta$ in equation~(\ref{eqVn}) is
introduced through the artificial friction force $\vec{f}_{\zeta}\equiv-\zeta n_{n}\vec{v}_{n}$ acting on the neutrons in order to follow the long-term quasi-stationary evolution \citep{hoyos2008magnetic,castillo2020twofluid}. This force replaces and mimics  
the short-term dynamics associated to the inertial terms in the Euler equations, which evolve the NS core towards a hydromagnetic quasi-equilibrium state, in which all the physical (i.~e., non-fictitious) forces acting on the fluid elements are close to balancing each other: $\vec f_B+\vec f_n+\vec f_c\approx 0$. This approximation is justified because the fluid in the NS core is expected to reach the quasi-equilibrium state with the magnetic field within a few Alfv\'en crossing times $\sim 1\,(10^{14}\,\text{G}/B)\, \text{s}$, and we are interested in the NS evolution on much longer time scales.

In equation~(\ref{eqVad}), $\gamma_{cn}$ is a rate coefficient that parametrizes the frictional drag forces due to collisions between charged particles and neutrons, and can be written as \citep{Yakovlev1991}
\begin{equation}\label{eqGammacn}
    \gamma_{cn}\approx 5.0\times 10^{-44}\left(\dfrac{T}{10^{9}\,\text{K}}\right)^2\left(\dfrac{\rho}{\rho_{\text{nuc}}}\right)^{-1/3}
    \dfrac{n_n}{n_{\text{nuc}}}
    \,\text{g cm}^{3}\text{s}^{-1},
\end{equation}
where $\rho$ is the mass density and $n_{\text{nuc}}=0.16\,\text{fm}^{-3}$ and $\rho_{\text{nuc}}=2.8\times 10^{14}\text{g\,cm}^{-3}$
are the nuclear saturation number density and mass density, respectively. Here, $T$ is the core temperature.

The velocity fields $\vec{v}_n$ and $\vec{v}_{ad}$ are also further constrained to satisfy the continuity equations at all times, namely,
\begin{gather}
    \nabla \cdot (n_{n}\vec{v}_{n})= + \Delta \Gamma,\label{eqCont1}\\
    \nabla \cdot \left(n_{c}\vec{v}_{c}\right)= -\Delta \Gamma,\label{eqCont2}
\end{gather}
where the time derivatives of the number density perturbations, $\delta n_{i}$ ($i=c,n$), have been neglected since a simple estimate gives $|(\partial\delta n_i/\partial t)/[\nabla \cdot (n_{i}\vec{v_{i}})]| \sim B^{2}/8\pi P\ll 1$. Here, $\Delta \Gamma$ is the net conversion rate per unit volume of charged particles to neutrons by non-equilibrium Urca processes (direct or modified), which in general has a non-linear dependence on the variable $\Delta \mu /\pi k_{B}T$ \citep{Reisenegger1995}, where $\Delta \mu \equiv \delta \mu_{c}-\delta \mu_{n}$ is the chemical imbalance and $k_{B}$ the Boltzmann constant. However, we will see that in the present problem the departure from chemical equilibrium remains modest, $|\Delta \mu| \lesssim \pi k_{B}T$, in the so-called ``sub-thermal'' regime or approximation \citep{Haensel2002}, in which $\Delta \Gamma$ is proportional to the chemical imbalance 
\begin{equation}\label{eqLambda}
    \Delta \Gamma = \lambda \Delta \mu,
\end{equation}
where $\lambda$ is a function that depends strongly on temperature and on the specific Urca processes involved. 

In what follows in this work, the microphysical input is obtained by using the HHJ equation of state (EoS) \citep{Heiselberg_1999}, for a NS with a total mass $M=1.4\,M_\odot$, radius $R=12.2\,\text{km}$, core radius $R_{core}=11.2\,\text{km}$, central mass density $\rho_0= 9.3\times 10^{14}\,\text{g}\,\text{cm}^{-3}$, central pressure $P_0 = 1.2\times 10^{35}\,\text{erg}\,\text{cm}^{-3}$,central number densities $n_{n0}=4.7\times 10^{38}\,\text{cm}^{-3}$ and $n_{c0}=4.2\times 10^{37}\,\text{cm}^{-3}$ for neutrons and charged particles, respectively, and central matrix elements $K_{cc,0}=4.5\times 10^{-42}\,\text{erg}\,\text{cm}^{3}$, $K_{nn,0}=1.3\times 10^{-42}\,\text{erg}\,\text{cm}^{3}$, and $K_{nc,0}= K_{cn,0} = 7.3\times 10^{-43}\,\text{erg}\,\text{cm}^{3}$. 

For this EoS, the direct Urca mechanism is allowed in the core only for stellar masses $M>1.83\,M_\odot$. Hence, we focus exclusively on the modified Urca process, for which $\lambda$ is given by \citep{gusakov2017evolution}
\begin{align}\label{eqLambdaMU}
    \lambda \approx 5.0\times 10^{33}\,\left(\dfrac{T}{10^{9}\,\text{K}}\right)^{6}\left(\dfrac{\rho}{\rho_{\text{nuc}}}\right)^{2/3}\text{erg}^{-1}\text{cm}^{-3}\text{s}^{-1}.
\end{align}


\subsection{``Strong-coupling'' regime} \label{secStC}

Since the coefficients $\lambda$ and $\gamma_{cn}$ depend on temperature, it is necessary to also follow the evolution of the temperature in order to obtain the evolution of the magnetic field. At any given time, and thus for a given magnetic field configuration and temperature, equations~(\ref{eqVn})-(\ref{eqVad}) and (\ref{eqCont1})-(\ref{eqCont2}) would need to be used to obtain the chemical potential perturbations and the velocities, which would then be used to take a step in time to calculate the new temperature and magnetic field configuration.

In order to simplify this process, we note that Urca reactions and ambipolar diffusion have  
strong (and opposite) dependencies on the temperature, namely $\lambda \propto T^{6}$ and $v_{ad}\propto\gamma_{cn}^{-1}\propto T^{-2}$ (see equations \ref{eqGammacn} and \ref{eqLambdaMU}). Due to this strong dependence, the moment when both mechanisms operate with equal efficiency implicitly defines a temperature, $T_{trans}$, that marks the transition from the strong-coupling to the weak-coupling regime, and whose derivation will be given in \S~\ref{sec:logtev}.
Here, we focus on the strong-coupling regime, which corresponds to high temperatures $T>T_{trans}$ and can be described by setting the ambipolar velocity to zero ($\gamma_{cn}\to\infty$).
Thus, $ \vec{v}_n = \vec{v}_{c} \equiv \vec{v}$, and the full set of equations to solve in axial symmetry becomes
\begin{gather}
    \dfrac{\partial \alpha}{\partial t}= -\vec{v}\cdot \Grad \alpha,\label{eqIndA}\\
     \dfrac{\partial \beta}{\partial t}=r^{2}\sin^{2} \theta \, \Grad \cdot \left[ \dfrac{\left(\vec{v}\times \vec{B}\right)\times \hat{\phi}}{r \sin \theta}\right],\label{eqIndB}\\
     \nabla \cdot (n_{n}\vec{v})= +\lambda\Delta \mu,\label{eqContSC1}\\
     \nabla \cdot (n_{c}\vec{v})= -\lambda\Delta \mu,\label{eqContSC2}\\
      \vec{v}= \dfrac{1}{\zeta n_{n}}\left(\dfrac{\Curl\vec{B}}{4\pi}\times \vec{B}-n_{n}\mu\Grad\chi_{n}-n_{c}\mu\Grad\chi_{c}\right),\label{eqVSC}\\
      \Delta \mu = \mu(\chi_c -\chi_n).\label{eqDmu}
\end{gather}
Equations~(\ref{eqContSC1})-(\ref{eqDmu}) must be solved first to obtain the chemical potential perturbations and the velocity for a given magnetic field configuration, and then the obtained velocity must be used to evolve the magnetic field through equations~(\ref{eqIndA}) and  (\ref{eqIndB}). For the first step, we replace 
the expression for the velocity (equation~\ref{eqVSC}) into the continuity equations (\ref{eqContSC1})-(\ref{eqContSC2}), 
obtaining
\begin{gather}
\Grad\cdot\left(n_{n}\mu\Grad \chi_{n}+n_{c}\mu\Grad \chi_{c}\right)+\zeta\lambda \mu \Delta \chi=\Grad \cdot \left(\dfrac{\Curl \vec{B}}{4\pi}\times \vec{B}\right)\label{eqMatrix1},\\
\Grad\cdot\left[\dfrac{n_{c}}{n_{n}}(n_n\mu\Grad\chi_{n}+n_c\mu\Grad\chi_{c})\right]-\zeta\lambda \mu\Delta\chi=\Grad \cdot \left(
\dfrac{n_{c}}{n_{n}}
\dfrac{\Curl \vec{B}}{4\pi}\times \vec{B}\right),\label{eqMatrix2}
\end{gather}
where $\Delta\chi\equiv\chi_c-\chi_n$. For a given magnetic field configuration (which determines the right-hand side of both equations) and with the additional constraint of baryon number conservation (equation~\ref{eqBaryon}), these can be solved numerically for $\chi_n$ and $\chi_c$ by a finite-element method. 
Then, it is easy to compute the velocity $\vec v$ through equation~(\ref{eqVSC}) and forward-step the magnetic field components through equations~(\ref{eqIndA}) and (\ref{eqIndB}), as illustrated in Fig.~\ref{fig1}.


\subsection{Boundary conditions}\label{sec:boundarycond}
In the work of \citet{Castillo2017,castillo2020twofluid}, 
it was assumed that the currents in the crust decay much faster than the typical evolution time scales in the core, so the crust was treated as a vacuum (a perfect resistor) whose magnetic field at any time is fully determined by the field in the core. This is unlikely to be satisfied in the strong-coupling regime because of the short time scales involved and because the crust is still partly liquid and only progressively freezes during this period of time (e.~g., \citealt{Aguilera2008}). 
Nevertheless, and for simplicity, we proceed in the same way as \citet{Castillo2017,castillo2020twofluid}, considering the external magnetic configuration as a current-free, poloidal magnetic field that is computed at all time-steps as a multipolar expansion and imposing continuity of the poloidal component at the crust-core interface, $r=R_{core}$. We note that this assumption should allow a faster evolution than the opposite extreme of a magnetic field frozen into the crust, and even so we will see that the observed evolution is very limited.

At the crust-core interface, we also assume that the radial component of the fluid velocity $\vec{v}$ vanishes. Thus, from equation (\ref{eqVSC}), we obtain 
\begin{equation}
   \left[f_{B}^{r}-n_{c}\mu \dfrac{\partial \chi_{c}}{\partial r}-n_{n}\mu \dfrac{\partial \chi_{n}}{\partial r}\right]_{r=R_{core}}=0.
\end{equation} 
On the other hand, in the strong-coupling regime, the radial velocity component can be determined solely from the continuity equations (\ref{eqContSC1})-(\ref{eqContSC2}), leading to the expression 
\begin{equation}\label{eqVr}
    v_{r}=\dfrac{\ell_{c}\lambda \Delta \mu}{n_c},
\end{equation}
where we introduced a length scale
\begin{equation}\label{eq:lc}
    \ell_{c}\equiv -[d\ln(n_{c}/n_{b})/dr]^{-1},
\end{equation}
with the baryon density given by $n_{b}=n_{c}+n_{n}$. 
As a consequence, at the crust-core interface, the chemical equilibrium condition is satisfied, i.~e.
\begin{equation}
    \chi_{c}(r=R_{core})=\chi_{n}(r=R_{core}).
\end{equation}

Further boundary conditions are related to the axial symmetry constraint. In order for the velocity field and the magnetic field to be continuous at the axis, one must impose $v_{\theta}=v_{\phi}=0$ and $B_{\theta}=B_{\phi}=0$ at $\theta=0$ and $\theta=\pi$. 
The latter conditions imply $\partial \alpha / \partial r (\theta=0,\pi) = \beta(\theta=0,\pi) = 0$. Hence, $\alpha$ must be a constant along the axis, whose value we set to zero, so that the magnetic flux enclosed by any circle around the axis at fixed $r$ and $\theta$ can be obtained as $2\pi\alpha(r,\theta)$.



\subsection{Time scales}\label{sec:timescales}

\subsubsection{Short-term evolution through artificial friction}
In \citet{castillo2020twofluid} and in the present work, in order to focus on the long-term evolution, the inertial terms in the equations of motion for neutrons and charged particles are neglected, and the short-term dynamics associated to the excitation and relaxation of sound, gravity, and Alfv\'en waves is mimicked by introducing an artificial friction force, $\vec{f_{\zeta}}=-\zeta n_n \vec v_n$, acting on the neutron fluid. 

In \citet{castillo2020twofluid}, which considered the weak-coupling regime, the aim was to start with an arbitrary initial magnetic field configuration, and then, by properly choosing $\zeta$, allow the bulk motions controlled by $\vec{f_{\zeta}}$ to reduce the net force imbalance on a fluid element (in parenthesis on the right-hand side of equation \ref{eqVn}) much more quickly than ambipolar diffusion (with relative velocity $\vec{v_{ad}}$) reduced the force imbalance on the charged-particle component (in parenthesis on the right-hand side of equation \ref{eqVad}).  
This implied that $\zeta$ had to fulfill $\zeta \ll \gamma_{cn}n_{c}$, while at the same time being large enough for a feasible numerical simulation. 
Thus, \citet{castillo2020twofluid} 
presented a detailed hierarchy of time scales that mimic the relaxation due to sound, gravity, and Alv\'en waves propagating through the NS core with artificial friction. The first two represent the typical time scales in which the density perturbations grow and the poloidal hydromagnetic quasi-equilibrium is reached. 

In the new scheme presented here, however, the time derivative terms in the continuity equations are neglected, which is valid for times longer than the gravity or buoyancy time scale ($t_{\zeta g}$, in the notation of \citealt{castillo2020twofluid}). This approach speeds up the simulation code and allows to consider realistic magnetic field strengths since the simulations now start after this buoyancy-like time scale, when the (non-fictitious) poloidal forces are already close to a hydromagnetic equilibrium ($\vec{f}^{\text{Pol}}_{B}+\vec f_n+\vec f_c\approx 0$). 
Nevertheless, from this new starting point, the toroidal Lorentz force, $\vec{f}^{\text{Tor}}_{B}$, is still finite and  balanced by fictitious friction. It must decay to zero by 
toroidal fluid motions, with velocity $v^{\text{Tor}} \sim B^{2}/(4 \pi \zeta n_n \ell_B)$ (see equation~\ref{eqVSC}), generated by the magnetic field itself,
which occurs on a longer, Alfv\'en-like time scale
\begin{equation}\label{eqTBz}
      t_{\zeta B}\sim \dfrac{\ell_{B}}{v^{\text{Tor}}}\sim \frac{4 \pi n_{n}\ell_{B}^{2}\zeta}{B^{2}},
      \end{equation}
where $\ell_{B}$ is the typical length-scale of the magnetic field. 

 \begin{table*}
\centering
\begin{tabular}{ |p{4.2cm}||p{2.5cm}||p{2.4cm}||p{3.5cm}|}
 \hline
   & \quad \quad \quad \quad \, Code units  \quad \\
 \hline
 Variable name & Notation   &  Normalization  & Value\vspace{0.1cm}\\
\hline
Magnetic field strength & \quad $B$ & $B_{init}$ &\,\,\,\,\,--------- \vspace{0.15cm}\\
Temperature &\quad $T$ & ------&\,\,\,\,\,---------  \vspace{0.15cm}\\
Distances & $\ell_c,\,\ell_B,\,r$ & $R_{core}$ & $11.2\,\text{km}$ \vspace{0.15cm} \\
Poloidal flux function &\quad $\alpha$ & $B_{init}R_{core}^{2}$ & \,$1.3\times 10^{29}B_{init,15}\,\text{G cm}^{2}$ \vspace{0.15cm}\\
Poloidal current function &\quad $\beta$ & $B_{init}R_{core}$ & \,$1.1\times 10^{24}B_{init,15}\,\text{G cm}$ \vspace{0.15cm}\\
Number densities & $n_j\,(j=n,c)$     & $n_{c0} \equiv n_c(r=0)$ & $4.2\times 10^{37}\,\text{cm}^{-3}$ \vspace{0.15cm}\\
Chemical potential &\quad $\mu$ & $\mu_0 \equiv \mu(r=0)$ & $1.9\times 10^{-3}\,\text{erg}$ \vspace{0.15cm} \\
Chemical perturbations & $\,\,\delta\mu_j$ & $\frac{B_{init}^{2}}{4\pi n_{c0}}$ & $1.9\times 10^{-9}B^{2}_{init,15}\,\text{erg}$ \vspace{0.15cm} \\
$\lambda$- function &\quad $\lambda$ & $\lambda_0 \equiv \lambda(r=0)$ & $1.1\times10^{34}\,T_{9}^{6}\,\text{erg}^{-1}\text{cm}^{-3}\text{s}^{-1}$ \vspace{0.15cm} \\
Artificial friction coefficient &\quad $\zeta$ &$ \frac{n_{c0}}{\lambda_0 R_{core}^{2}}$ & $3.1\times 10^{-9}\,T_{9}^{-6}\,\text{g}\,\text{s}^{-1}$  \vspace{0.15cm}\\
Time &\quad $t$ & $ \frac{4\pi n^{2}_{c0}}{\lambda_0 B_{init}^{2}}$ & $6.5 \times 10^{4} B^{-2}_{init,15}T_{9}^{-6}\,\text{yr}$ \vspace{0.15cm}\\
Magnetic energy &\quad $E_{B}$ & $ \frac{B_{init}^{2}R_{core}^{3}}{6}$ & $2.3 \times 10^{47}B_{init,15}^{2}\,\text{erg}$  \vspace{0.15cm}\\
Luminosities &$L_{H\nu}\,,L_{ad}\,, L_{\zeta}\,,L_{\nu}^{*}$ & $ \frac{\lambda_0 B_{init}^{4}R_{core}^{3}}{24\pi n^{2}_{c0}}$ & $1.1 \times 10^{35}B_{init,15}^{4}T_{9}^{6}\,\text{erg}\,\text{s}^{-1}$  \\

 \hline
\end{tabular}
 \caption{Summary of the code units, showing the normalization of the different variables, their notations and physical values obtained from the HHJ EoS, for a NS that has a total radius $R=12.2\,\text{km}$, a core radius $R_{core}=11.2\,\text{km}$, and a total mass $M=1.4\,\textup{M}_\odot$. The definitions of $L_{H\nu}\,,L_{ad}\,, L_{\zeta}$, and $L_{\nu}^{*}$ are given in \S~\ref{sec:MagneticDiss}. Here, the free parameters are the initial rms magnetic field strength, $B_{init}$, and the temperature, $T$, which are needed to recover the physical units from the simulation output. The notation $B_{init,15}$ means $B_{init}/10^{15}\,\text{G}$. }
 \label{table2}
\end{table*}

\subsubsection{Long-term evolution through Urca reactions }\label{sec:logtev}
As we have previously discussed, the short-term dynamics, mimicked by the fictitious friction, evolves the NS core towards an almost complete balance of forces, leaving it 
out of chemical equilibrium. The excess of chemical energy can only be dissipated by Urca reactions, which in turn locally change the degeneracy pressure forces, so that the magnetic field must rearrange itself, moving particles in such a way that a new hydrostatic equilibrium is reached. As we shall explain in detail in this section, this process changes the typical time scale at which the core reaches chemical equilibrium, and the time scale for the decay of $\Delta\mu$ can be much longer
when considering the effect of the magnetic field dynamics than in the absence of a magnetic field \citep{reisenegger2009stable}. 
We emphasize this because various authors (e.~g., \citealt{DuncanThompson1996,lander221protoNS}) have assumed that, since the initial NS temperature is so high, the stellar matter can be regarded to be in chemical equilibrium and thus to be described as a barotropic fluid.
As we shall show in \S~\ref{sec:GSequilibrium}, this assumption imposes strong, but unrealistic, constraints on the initial magnetic field configuration. 

By studying the time derivative of $\Delta \mu$, the typical time scale of the long-term magneto-chemical evolution can be derived. Using the continuity equations (\ref{eqContSC1}) and (\ref{eqContSC2}) (but now retaining the time-derivatives of the number densities on their left-hand sides), as well as 
equation (\ref{eqPerturb}), one obtains
\begin{align}\label{eqDtDmu}
\dfrac{\partial }{\partial t}\Delta \mu &  = -\lambda(K_{cc}+K_{nn}-2K_{nc})\Delta \mu 
+(K_{nn}-K_{nc})\nabla \cdot(n_{n} \vec{v})\\\nonumber
&+(K_{nc}- K_{cc}) \nabla \cdot (n_{c}\vec{v}).
\end{align}
This expression shows that, if the magnetic field suddenly disappeared and there were no fluid motions, 
$\Delta \mu$ would decrease exponentially, reaching 
chemical equilibrium on a
time scale 
\begin{equation}\label{eqtlambda}
  t_{\lambda} \sim\dfrac{1}{\lambda K_{cc}} 
  \sim 0.6\,\left(\dfrac{T}{10^{9}\,\text{K}}\right)^{-6}\,\text{yr},
\end{equation}
where we used $K_{cc}\gtrsim K_{nn},\vert K_{nc}\vert $, and evaluated $K_{cc}$ and $\lambda$ at the center ($r=0$) to give the numerical estimate.

\begin{figure*}
    \centering
    \includegraphics[width=18.1cm]{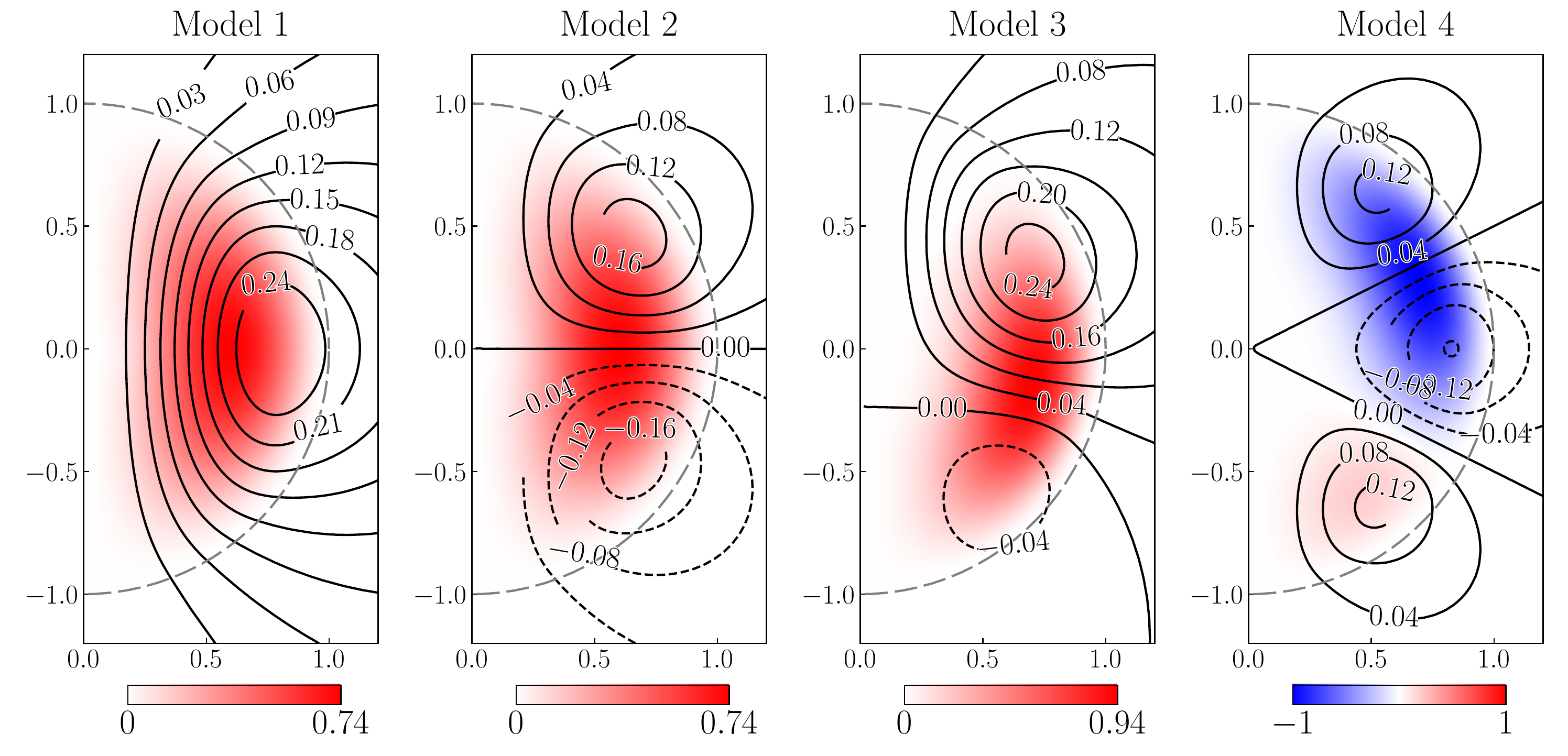}
    \caption{Initial magnetic field configurations used in our simulations given by the potentials listed in Table~\ref{table1}. The lines represent the poloidal magnetic field, labeled by the magnitude of $\alpha_{init}$, and in colors the poloidal current function $\beta_{init}$. These variables are plotted in code units, see Table.~\ref{table2}.}
    \label{fig2}
\end{figure*}
\begin{table*}
\centering
\begin{tabular}{ |p{.5cm}||p{4cm}||p{2cm}|p{1cm}|p{5.5cm}|p{1cm}| }
 \hline
 Model & \quad \quad \quad \quad $\alpha_{init}(r,\theta)/(\sqrt{0.6} \alpha_{0})$ &\quad $a_{1}$, $a_{2}$ & $\alpha_{0}$ & \quad  $\beta_{init}(r,\theta)/(\sqrt{0.4} \beta_{0})$ & \quad $\beta_{0}$\vspace{0.15cm}\\
 \hline
 \quad 1    & \quad \quad \quad \quad \quad \quad \quad  $ \alpha_{1} $ & $-6/5$, $3/7$ & 1.34& \quad  $r^{3}(1-r)^{2}\sin^{2}\theta$ & 33.98\vspace{0.15cm}\\
 \quad 2   & \quad \quad \quad \quad \quad \quad \quad $\alpha_{2}$ & $-10/7$, $5/9$ &   1.24 &  \quad $r^{3}(1-r)^{2}\sin^{2}\theta$ & 33.98\vspace{0.15cm}\\
 \quad 3&  \quad \quad \quad \quad \quad $0.3\alpha_{1}+0.7\alpha_{2}$  & \quad --------- &  \quad 1  &\quad$r^{5}(1-r)^{2}\sin^{2}\theta \sin\left(\theta-\frac{\pi}{5}\right)$& 112.54\vspace{0.15cm}\\
 \quad 4 & \quad \quad \quad \quad \quad \quad \quad$\alpha_{3}$ &   $-14/9$, $7/11$ & 1.27& $r^{6}(1-r)^{2}\sin^{2}\theta \left[\sin\left(\theta-\frac{\pi}{7}\right)-\sin\left(\theta-\frac{\pi}{5}\right)\right]$ & 1158.77\\
 \hline
\end{tabular}
 \caption{Parameters for the initial magnetic field configurations used in our simulations. The normalization constants $\alpha_{0}$ and $\beta_{0}$ are fixed by the condition $\langle \vec{B}^{\,\text{Pol}} \rangle =1$, and $\langle \vec{B}^{\,\text{Tor}} \rangle =1$, respectively, where $\langle . \rangle$ denotes an rms average over the stellar core. Here, we choose the poloidal potentials as $\alpha_{\ell}\equiv -r^{\ell+1}(1+a_{1}r^{2}+a_{2}r^{4})P^{1}_{\ell}(\cos \theta)$, where $P^{m}_{\ell}(\cos \theta)$ is the associated Legendre polynomial of order $\ell$ with azimuthal index $m$; and the coefficients, $a_{1}$ and $a_{2}$, are determined so that the azimuthal current vanishes at the stellar surface, $J_{\phi}(r=R,\theta)=0$, and $\alpha$ matches the core magnetic field with the external multipolar expansion corresponding to a current-free magnetic configuration (see \S~\ref{sec:boundarycond}) \citep{Armaza_2015}. The constants $\sqrt{0.6}$ and $\sqrt{0.4}$, multiply the potentials in order to set 60\% and 40\% of initial poloidal and toroidal magnetic energy, respectively. These models are shown in Fig.~\ref{fig2}.} 
 \label{table1}
\end{table*}

However, when the magnetic and velocity fields are present, there are fluid motions that try to keep the force balance and largely compensate the effect of the Urca reactions. 
The time scale of this coupled magnetic and chemical (magneto-chemical) evolution follows from the continuity equations, which yield a
characteristic fluid velocity that can be estimated from equation (\ref{eqVr}) as
    $v\sim \lambda \vert \Delta\mu \vert \ell_{c}/n_{c}.$ 
On the other hand, considering zero artificial friction in 
equation (\ref{eqVn}), the degeneracy pressure forces plus gravity balance the poloidal Lorentz force, $\vec{f}_{B}^{\text{Pol}}+\vec{f}_{c}+\vec{f}_{n}=0$. Then, assuming $\vert \vec{f}_{B}^{\text{Pol}} \vert \sim \vert \vec{f}_{c}\vert \sim \vert \vec{f}_{n}\vert$ (as we will see in \S~\ref{sec:magchem}, this assumption is not strictly correct, but here we only aim to provide a rough estimate of $t_{\lambda B}$), so that $n_c \vert \Delta \mu \vert \sim B^{2}/4\pi$, the characteristic time scale of this process is given by
\begin{align}\label{eqTBlamb}
   t_{\lambda B} & \sim \dfrac{\ell_{B}}{v}\sim \dfrac{4\pi n^{2}_{c}}{\lambda B^{2}}
   \dfrac{\ell_{B}}{\ell_{c}}, 
   \\\nonumber
   & \sim 2.0\times10^{6}\,
   \dfrac{\ell_{B}}{\ell_{c}}
   \,\left(\dfrac{B}{10^{15}\text{G}}\right)^{-2} \left(\dfrac{T}{10^{9}\text{K}}\right)^{-6}\,\left(\dfrac{n_c}{n_{\text{nuc}}}\right)^{2}\,\left(\dfrac{\rho}{\rho_{\text{nuc}}}\right)^{-2/3}\mathrm{yr}
\end{align}
(see also \citealt{DuncanThompson1996,Reisenegger2005,reisenegger2009stable}, but note that the first of these references incorrectly assigns this time scale to the low-temperature, ``weak-coupling'' regime).

We note that the ratio between the latter two time scales is roughly 
\begin{equation}\label{eqTlambTlambB}
   \dfrac{t_{\lambda}}{t_{\lambda B}} \sim 
   \dfrac{B^{2}}{4\pi n^{2}_c K_{cc}} \dfrac{\ell_c}{\ell_B} \sim \dfrac{B^{2}}{8\pi P}\dfrac{n_b}{n_c}\dfrac{\ell_c}{\ell_B}
   \sim \left(\dfrac{B}{3 \times 10^{17}\,\text{G}}\right)^{2} \dfrac{\ell_c}{\ell_B},
\end{equation}
where we used $n_c^{2} K_{cc} \sim P n_c/n_b$ and again evaluated at the center ($r=0$) to give the numerical estimate. It can be seen that,  
even for magnetar-strength magnetic fields, the time scale for the combined magneto-chemical evolution is expected to be much longer than the naive estimate $t_\lambda$. 
 
Here, we have consistently distinguished between the length scales $\ell_c$ and $\ell_B$ in equation (\ref{eqTBlamb}) and (\ref{eqTlambTlambB}) because, for an arbitrary initial magnetic field configuration, $\ell_c$ is typically larger than $\ell_B$. This stems from the fact that the radial profiles of the densities, $n_c(r)$ and $n_b(r)$ are not identical, but quite similar, smoothly decreasing over a lengthscale $\ell_c\sim R_{core}$ (see equation~\ref{eq:lc}), while $\ell_B$ is generally only a fraction of $R_{core}$.

In order to identify the transition temperature to the weak-coupling regime, we set $t_{\lambda B}$, evaluated at $r=0$, equal to the ambipolar diffusion time, as given by \citet{castillo2020twofluid}, and also evaluated at $r=0$,
\begin{align}
    t_{ad} &\sim 0.1 \dfrac{4\pi \gamma_{cn}n_n n_c \ell_B^{2}}{B^{2}},\\\nonumber 
     & \sim 2\times 10^{5}\, \left(\dfrac{B}{10^{15}\,\text{G}}\right)^{-2}\,\left(\dfrac{T}{10^{9}\,\text{K}}\right)^{2}\left(\dfrac{\ell_B}{2.8\,\text{km}}\right)^{2}\,\text{yr},
\end{align}
where the correction factor $0.1$ was included in consistency with the findings of \cite{ofegeim2018}
and \cite{castillo2020twofluid}. Estimating $\ell_B \approx R_{core}/4= 2.8\,\text{km}$ and $\ell_c\approx 4\ell_B\approx R_{core}$, this yields the transition temperature
\begin{equation} \label{eq:Ttrans}
 T_{trans}\approx 5\times 10^{8}\,\text{K}.
\end{equation}

Lastly, as in the weak-coupling regime, the value of $\zeta$ is not arbitrary. It must be set so as to separate the short (unphysical) and long (physical) time scales, $t_{\zeta B}\ll t_{\lambda B}$, which requires 
\begin{equation}\label{eqCondZeta}
    \zeta\ll\frac{n_c^2}{\lambda n_n\ell_B\ell_c}.
\end{equation}
Now, with the aid of equation (\ref{eqVr}), the radial component of the fictitious force can be written
as
\begin{equation}\label{eqRadalV}
    f_{\zeta,r}=-\dfrac{n_n}{n_c}\zeta\lambda\ell_c\mu\Delta\chi,
\end{equation}
for which equation~(\ref{eqCondZeta}) also implies $|f_{\zeta,r}|\ll|f_{c,r}|,|f_{n,r}|$. 

Thus, the condition in equation~(\ref{eqCondZeta}) not only guarantees a clear  
separation of the short and long time scales, but also consistently implies that, at all times, the poloidal component of the fictitious force is much smaller than
the non-fictitious poloidal forces, which therefore are close to balancing each other. 


\section{Evolution at constant temperature}\label{results_contantT}

The dimensionless code units are summarized in Table~\ref{table2}. These units have been selected to ensure that the system of equations (\ref{eqIndA})-(\ref{eqDmu}) becomes independent of the magnetic field strength and temperature. 

As the time scales $t_{\zeta B}$, $t_\lambda$, and $t_{\lambda B}$ depend on the number densities and other radial functions, hereafter, we use as references for them the expressions
\begin{equation}
    t_{\zeta B}\equiv \dfrac{4\pi n_{n0} \ell_B^{2} \zeta}{ B_{init}^{2}},
\end{equation}

\begin{equation}\label{eqtlambda_ref}
  t_{\lambda} \equiv\dfrac{1}{\lambda_0 K_{cc}(r=0)},
\end{equation}
and
\begin{equation}\label{eqtlambdaB_ref}
    t_{\lambda B}\equiv 
    \dfrac{4\pi n_{c0}^{2} \ell_B}{\lambda_0 B_{init}^{2}\ell_c}.
\end{equation}
In dimensionless code units, they read as
\begin{align}\label{eqtzBdiless}
t_{\zeta B} & = 
\frac{n_{n0}}{n_{c0}}
\left(\frac{\ell_B}{R_{core}}\right)^{2} \zeta = 0.684 \zeta,
\end{align}
\begin{align}
t_\lambda & = \frac{B_{init}^2}{4\pi n_{c0}^2 K_{cc}(r=0)} = \left(\frac{B_{init}}{3\times 10^{17}\,\mathrm{G}}\right)^2, \label{eqtlamdiless}
\end{align}
\begin{align}
t_{\lambda B} & = \frac{\ell_B}{\ell_c}=0.25,\label{eqtlamBdiless}
\end{align}
where, as before, we estimated $\ell_B /R_{core} \approx 1/4$, and $\ell_c\approx 4\ell_B\approx R_{core}$. 
We remark that in the dimensionless units used to express $\zeta$ (see Table.~\ref{table2}), it must satisfy the condition $\zeta \ll 4 n_{c0}/n_{n0} \approx0.4$
in order to maintain consistency with  equation~(\ref{eqCondZeta}). We also note that the time $t_\lambda$, which in code units becomes proportional to $B_{init}^2$ and very small for realistic values of this quantity, does not enter the evolution of the magnetic field at constant temperature, but will become relevant when analyzing the evolution with variable temperature in \S~\ref{sec:passive cooling}.

As an initial approach to the problem, we used the scheme explained above to evolve the set of equations (\ref{eqIndA})-(\ref{eqDmu}) \emph{at constant temperature} in order to study how the NS core reaches chemical equilibrium on the expected time scale. Also, here and hereafter, as initial condition for all the simulations we set the magnetic energy 
in the NS core to be 60\% poloidal and 40\% toroidal. 
\begin{figure}
    \includegraphics[width=8.5cm]{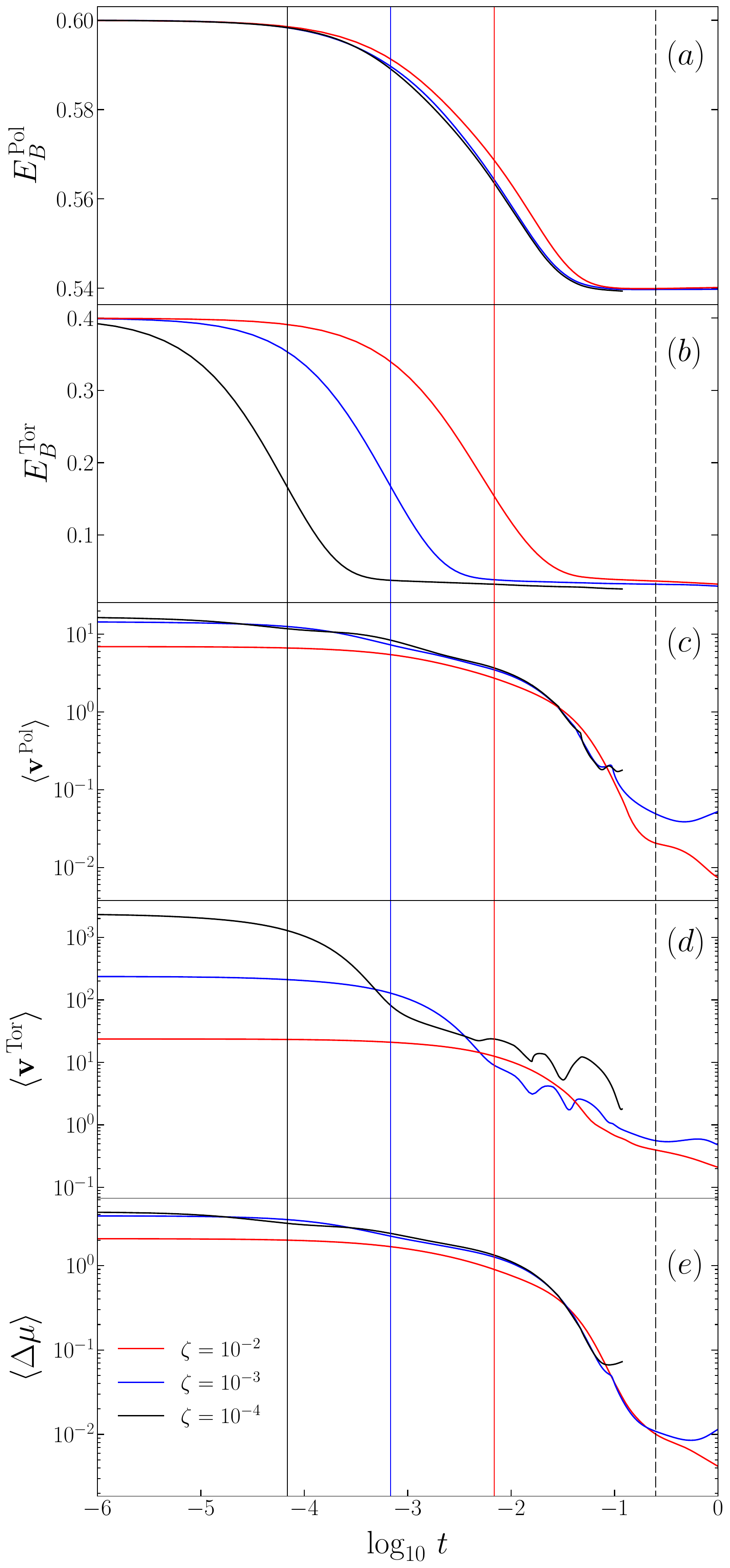}
    \caption{Comparison of the evolution of different simulations of the evolution at constant temperature using the same initial condition, Model 2 in Table~\ref{table1}, for three different values of $\zeta =10^{-2}$ (red), $10^{-3}$ (blue), and $10^{-4}$ (black),
    corresponding to the 
    ratios $t_{\lambda B}/t_{\zeta B}=37, 370,$ and $3700,$ 
    respectively. All quantities are given in code units, as defined in Table~\ref{table2}. The vertical lines, from left to right, show the values of the time scales $t_{\zeta B}$, with their respective colors in solid for the different $\zeta$ of each simulation, and $t_{\lambda B}$ (dashed). The panels show the time evolution of the poloidal fraction of the NS core magnetic energy in (a); the toroidal fraction of magnetic energy in (b); the root mean square (rms) averages over the volume of the core, $\langle\,\rangle$, of the poloidal and toroidal components of the fluid velocity in (c) and (d), respectively; and of the chemical imbalance, $\Delta\mu$ in (e).}
    \label{fig3}
\end{figure}

\subsection{Dependence on artificial friction}\label{sec:artificialfriction}

In order to separate the time scales in equations~(\ref{eqtzBdiless}) and (\ref{eqtlamBdiless}), 
we must set $\zeta \ll 1$ in dimensionless code units, which is equivalent to equation~(\ref{eqCondZeta}) in physical units. In order to make this condition more precise, we 
note that the long-term evolution must remain independent of the value of $\zeta$.  
We tested three different values, namely $\zeta=10^{-2},\,10^{-3}$ and $10^{-4}$, which yield the time-scale ratios $t_{\lambda B}/t_{\zeta B} = 37$, $370$, and $3700$, respectively. The results are summarized in Fig.~\ref{fig3}. 

In panels (a), (c), and (e), we see that the curves for $\zeta=10^{-3}$ and $\zeta=10^{-4}$ are almost identical from the beginning of the simulation, as expected for small enough $\zeta$, since this causes the non-fictitious poloidal forces to be near hydromagnetic quasi-equilibrium from the very beginning, as discussed in \S~\ref{sec:logtev}. More specifically, the good agreement between these curves can be interpreted in the sense that, for small enough $\zeta$, the poloidal force balance (the poloidal component of equation~\ref{eqVSC}) determines the chemical imbalances, but not the poloidal velocity field $\vec{v}^{\text{Pol}}$,
which is determined from the chemical imbalances by the continuity equations~(\ref{eqContSC1}) and (\ref{eqContSC2}). 
This good agreement is not seen for $\zeta = 10^{-2}$,
suggesting that this value is not small enough for the fictitious force to be ignored in the force balance.

For the toroidal variables,
the results are very
different. Panel (b) shows that, during the early stages ($t\lesssim t_{\zeta B}$), the toroidal magnetic energy decreases by the same amount for all values of $\zeta$, but on the different time scales $t_{\zeta B}$ associated to each value of $\zeta$. On the other hand, the toroidal velocity initially scales $\propto\zeta^{-1}$, and hence presents a more prominent evolution for the smaller values of $\zeta$. This happens because, for an arbitrary magnetic field configuration, the toroidal Lorentz force is only balanced by the corresponding component of the fictitious friction force. Thus, the magnetic field has to rearrange itself by moving the fluid in the toroidal direction with velocity $\vec{v}^{\text{Tor}}=\vec{f^{\text{Tor}}_{B}}/(\zeta n_n)$, until the toroidal hydromagnetic quasi-equilibrium is established on a typical time scale $\sim t_{\zeta B}$.

In panels (c), (d), and (e), we see that the velocity components and chemical imbalances strongly decrease towards late times, signaling an approach to hydromagnetic and chemical equilibrium. At this point, these variables become dominated by numerical noise (to a large extent due to the near-cancellation of forces in equation~\ref{eqVSC}), causing discrepancies between the curves for  different values of $\zeta$. Nevertheless, given the otherwise good agreement of the results with $\zeta = 10^{-3}$ and $10^{-4}$, we will hereafter use the value $\zeta = 10^{-3}$, which causes an acceptable separation of the time scales and is numerically cheaper than $\zeta = 10^{-4}$. 

\begin{figure}
    \centering
    \includegraphics[width=9.6cm]{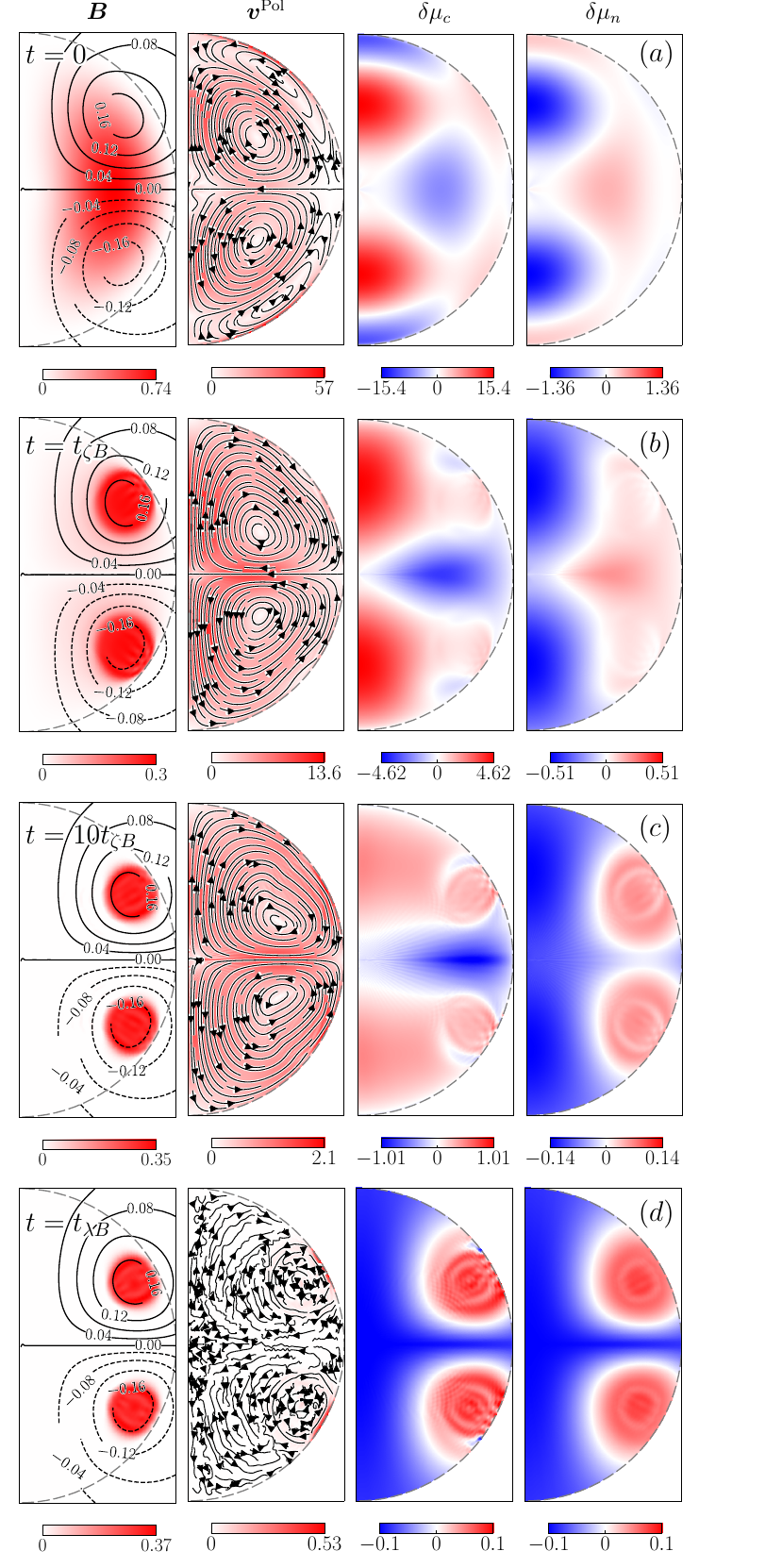}
    \caption{Magneto-chemical evolution at constant temperature for Model 2 with $\zeta=10^{-3}$, corresponding to $t_{\lambda B}/t_{\zeta B}=370$.  
    All panels are meridional cross-sections of the star. Rows (a), (b), (c), (d) correspond to different times, $t=0,t_{\zeta B},10\,t_{\zeta B}$, and $\,t_{\lambda B}$, respectively. In each row, the four panels display, from left to right: the magnetic field configuration, where lines represent the poloidal magnetic field, labeled by the magnitude of $\alpha$, and colors representd the poloidal current function $\beta$; the poloidal component of the velocity field, $\vec{v}$; the chemical potential perturbation of the charged particles; 
    and the chemical potential perturbation of the neutrons. 
   All quantities are given in code units, as defined in Table~\ref{table2}.}
    \label{fig4}
\end{figure}

\begin{figure}
    \centering
    \includegraphics[width=9.6cm]{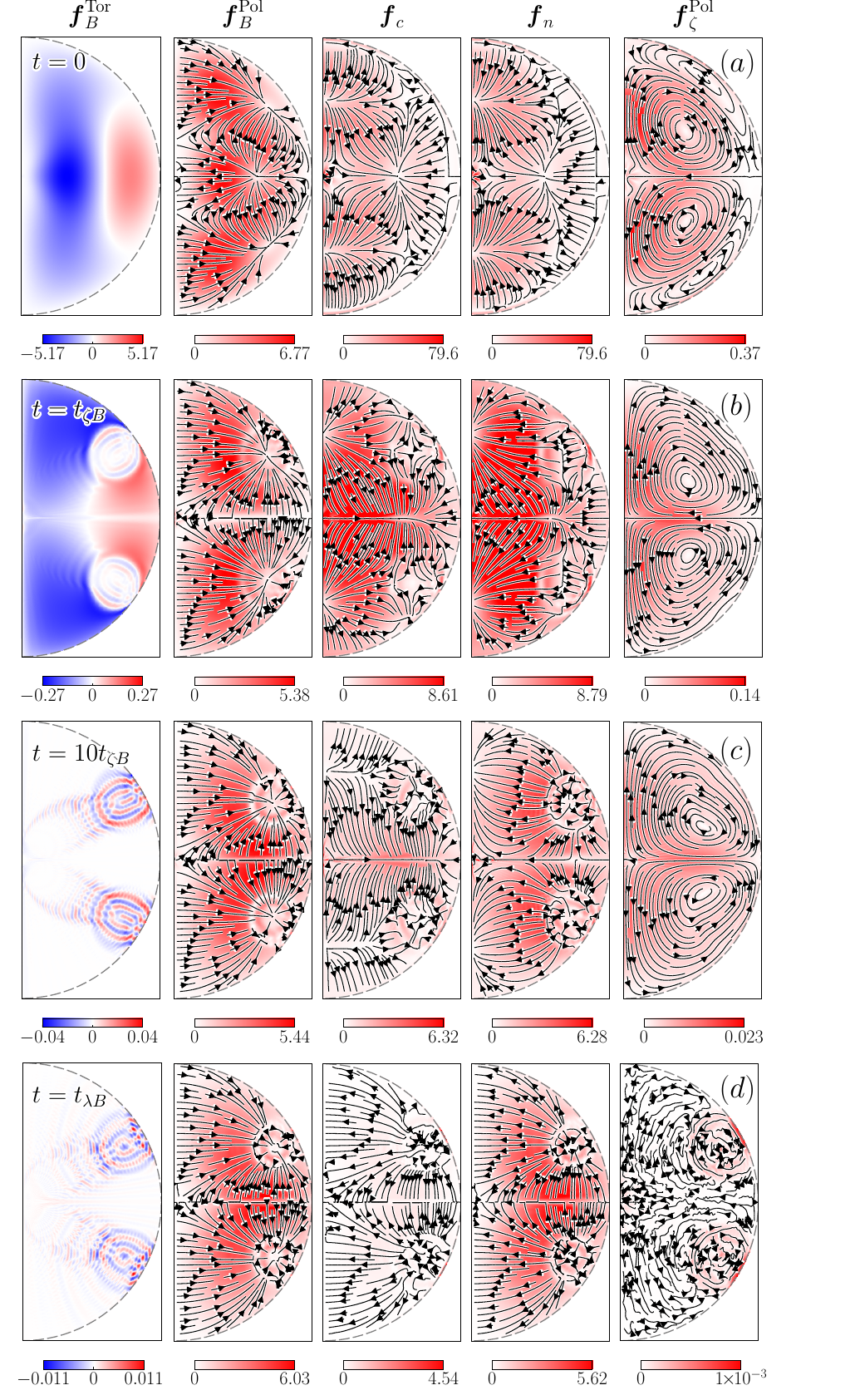}
    \caption{Evolution of the force densities 
    for the same simulation presented in Figs.~\ref{fig4} and
    \ref{fig6}. Rows (a), (b), (c), and (d) show snapshots at different times, $t=0,$ $t_{\zeta B},$ $10t_{\zeta B},$ and $t_{\lambda B}$, for the different force densities: From left to right, the toroidal Lorentz force $f_B^{\text{Tor}}$, the poloidal Lorentz force $\vec f_B^{\text{Pol}}$, the charged particle force $\vec f_c$, the neutron force $\vec f_n$, and the poloidal artificial friction force $\vec f_\zeta^{\text{Pol}}$. All quantities are given in code units, as defined in Table~\ref{table2}.} 
    \label{fig5}
\end{figure}

\subsection{Magneto-chemical evolution}\label{sec:magchem}
In this section, we present and discuss the results of our simulations for the ``magneto-chemical'' evolution at constant temperature, again using the results of Model 2 to describe some generic aspects of our simulations. The results for the other models (see Table~\ref{table1} and Fig.~\ref{fig2}) are analyzed later in \S~\ref{sec:GSequilibrium}. In any case, we have checked that the results for the magneto-chemical evolution are qualitatively the same for all the other models.

\begin{figure}
    \centering
    \includegraphics[width=8.5cm]{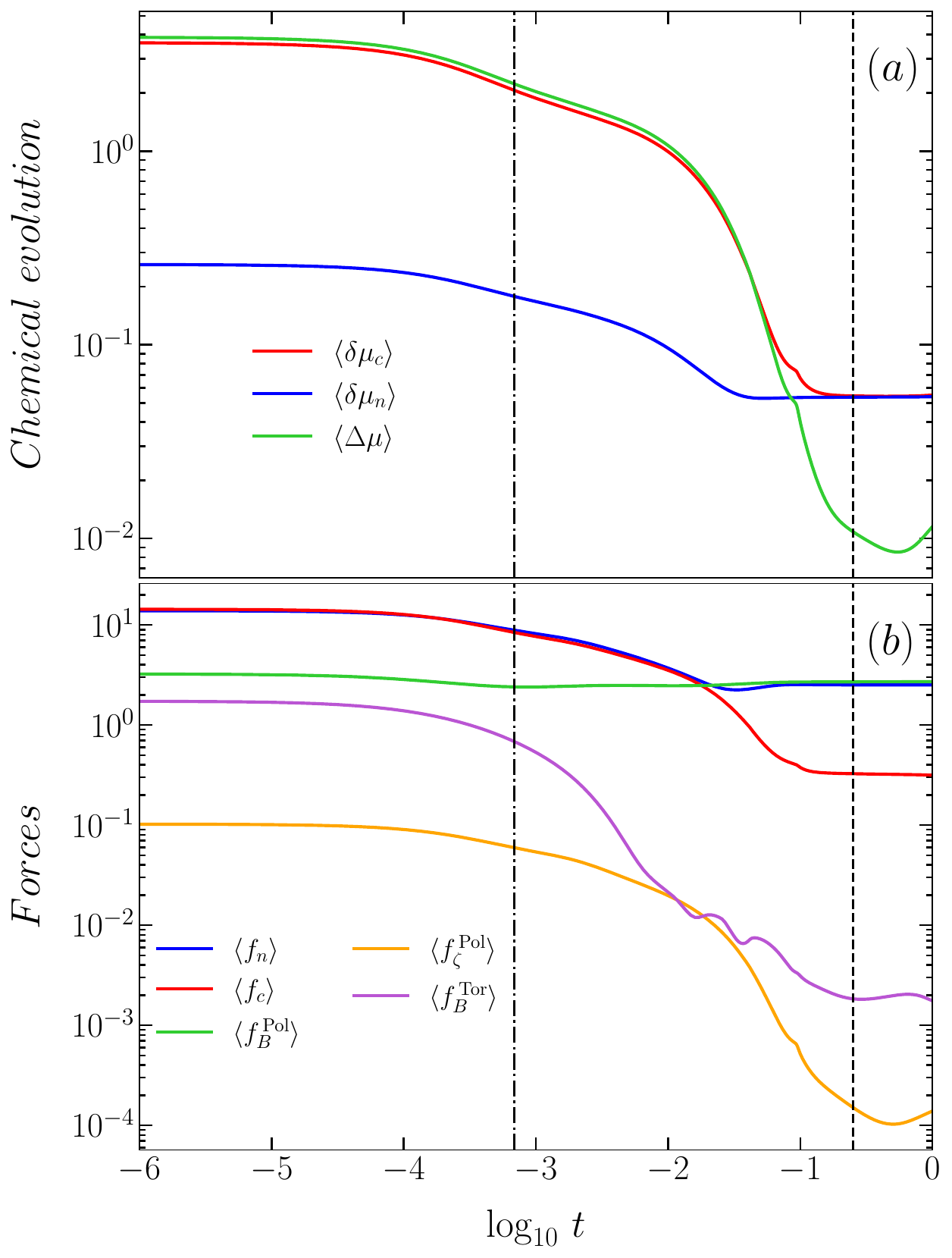}
    \caption{For the same simulation as in Figs.~\ref{fig4} and \ref{fig5}, panel (a) shows the rms values of the chemical potential perturbations $\delta \mu_{c}$ (blue), $\delta \mu_{n}$ (red), and $\Delta\mu$ (green); and panel (b) shows the rms values of the different forces: the poloidal Lorentz force $\vec f_B^\mathrm{Pol}$ (green), the charged-particle force $\vec f_c=-n_c\mu\Grad\chi_c$ (red), the neutron force $\vec f_n=-n_n\mu\Grad\chi_n$ (blue), the toroidal Lorentz force $\vec f_B^\mathrm{Tor}$ (purple), which is equal to minus the toroidal artificial friction force $-\vec{f_\zeta}^\mathrm{Tor}$, and the poloidal artificial friction force $\vec f_\zeta^\mathrm{Pol}$ (orange). 
    The vertical lines correspond, from left to right, to the time scale $t_{\zeta B}$ (dashed-dotted), and $t_{\lambda B}$ (dashed), respectively. Here, $\langle \rangle$ denotes the root mean square (rms) average over the volume of the core. All quantities are given in code units, as defined in Table~\ref{table2}.} 
    \label{fig6}
\end{figure}

The magneto-chemical evolution of Model 2 is presented in Figs.~\ref{fig4} and \ref{fig5}, which show the spatial dependence of different variables at selected times, and in Fig.~\ref{fig6}, which shows the time evolution of integrated or rms-averaged quantities. The initial state is clearly not a hydromagnetic quasi-equilibrium, as seen by the fact that the toroidal magnetic field is finite on open poloidal field lines, causing a substantial toroidal Lorentz force, which is balanced only by the corresponding component of the fictitious friction. At the same time, the chemical potential perturbations of the neutrons and charged particles are very different from each other (both in magnitude and in their spatial dependence), implying a substantial departure from chemical equilibrium. 

Nevertheless, the initial poloidal component of the artificial friction force, $\vec{f_{\zeta}}^{\text{Pol}}$, is already much weaker than the other poloidal forces, i.~e., the simulations start with a very well-established \emph{poloidal} quasi-equilibrium. We also note the solenoidal appearance of the poloidal velocity field, a consequence of the condition obtained from adding equations~(\ref{eqContSC1}) and (\ref{eqContSC2}), $\nabla\cdot(n_b\vec v)=0$, where $n_b\equiv n_n+n_c$ is the baryon number density.

Figs.~\ref{fig5}(a) and \ref{fig6}(b) show that initially $\vec f_{c} \approx -\vec f_{n}$, i.~e., the fluid forces (pressure plus gravity) on the charged particles and the neutrons roughly balance each other and are substantially stronger (by a factor $\sim 10$) than the poloidal Lorentz force $\vec f_B^{\text{Pol}}$. 
This behavior is at first sight quite surprising, because the chemical potential perturbations causing $\vec f_c$ and $\vec f_n$ are due to $\vec f_B^{\text{Pol}}$. However, it can be understood as follows. Since $\vec{f_{\zeta}}^{\text{Pol}}$ is from the beginning smaller than the other poloidal forces, the poloidal force balance equation can be approximated as
\begin{equation}\label{eqForcebalance}
    n_{c}\mu \Grad \chi_{c}+n_{n}\mu\Grad \chi_{n}\approx \vec{f}_{B}^{\text{Pol}}.
\end{equation}
Dividing this equation by $n_{c}\mu$ and taking the curl, one obtains 
\begin{equation}
   f_{n,\theta}=-\dfrac{\mu n_{n}}{r} \dfrac{\partial \chi_{n}}{\partial \theta } =
   \frac{n_c\mu\hat{\phi} \cdot \Curl[\vec{f}_{B}^{\text{Pol}}/(n_c\mu)]}{d\ln(n_c/n_n)/dr}.
\end{equation}
Repeating the same procedure, but dividing by $n_{n}\mu$ instead, one also obtains 
\begin{equation}
 f_{c,\theta}=-\dfrac{\mu n_{c}}{r} \dfrac{\partial \chi_{c}}{\partial \theta } =
   -\frac{n_n\mu\hat{\phi} \cdot \Curl[\vec{f}_{B}^{\text{Pol}}/(n_n\mu)]}{d\ln(n_c/n_n)/dr}.
\end{equation}
Thus, considering that $\vec{f}_{B}^{\text{Pol}}$ varies on a length scale $\ell_B$, whereas $n_c$ and $n_n$ vary on a larger length scale $\ell_c$, it becomes clear that $f_{c,\theta} \approx -f_{n,\theta}$ and $|f_{c,\theta}| \approx |f_{n,\theta}|\sim(\ell_c/\ell_B)|f_{B,\theta}|$.

As discussed in the previous subsection, a full hydromagnetic quasi-equilibrium is approached around the Alfv\'en-like time scale $\sim t_{\zeta B}$, on which the open magnetic field lines unwind, eliminating the toroidal component, except in the regions of closed poloidal field lines, and thus reaching the ``twisted-torus'' configurations expected in axially symmetric hydromagnetic equilibria \citep{Prendergast1956, braithwaite2004fossil}. 

On the longer time scale $t_{\lambda B}$, Urca reactions become effective, reducing the magnitude of the chemical imbalance $\Delta \mu$ (see Fig.~\ref{fig6}[a]) by changing the chemical potential perturbations $\delta\mu_i$ in order to equate them (Fig.~\ref{fig4}[d]), in that way modifying the local poloidal force balance. Therefore, the magnetic field slowly rearranges itself, so that the evolution proceeds through a continuum of consecutive hydromagnetic quasi-equilibria. 

As the NS core approaches the final chemical equilibrium state, the chemical potential perturbations become equal, thus $\delta \mu_c=\delta \mu_n$ ($\chi_c=\chi_n\equiv\chi$) and $\vec f_n=-n_n\mu\nabla\chi=(n_n/n_c)\vec f_c$. Therefore, $f_n$ is larger than $f_c$ by a factor $n_n/n_c\sim 10$, and the Lorentz force becomes mostly balanced by the neutron force, $\vec f_B\approx -\vec f_n$ (Figs.~\ref{fig5}[d], \ref{fig6}[b]).

When the NS core approaches
the final magneto-chemical equilibrium state, the velocity decreases substantially
in comparison to its initial values, while the chemical potential perturbations become roughly equal. This can be seen in Fig.~\ref{fig4} where, at the final snapshot ($t=\,t_{\lambda B}$), the velocity field has decreased by two orders of magnitude with respect to the initial state and $\delta \mu_c\approx \delta \mu_n$. The strange structures seen in Fig.~\ref{fig4}(d), second column, are caused merely by numerical noise due to truncation error in the quasi-cancellation of the forces, suggesting that the final magneto-chemical equilibrium has 
been reached (see also  
Fig.~\ref{fig6} [a] and [b]).

\subsection{Grad-Shafranov equilibria}\label{sec:GSequilibrium}
\begin{figure*}
    \centering
    \includegraphics[width=17cm]{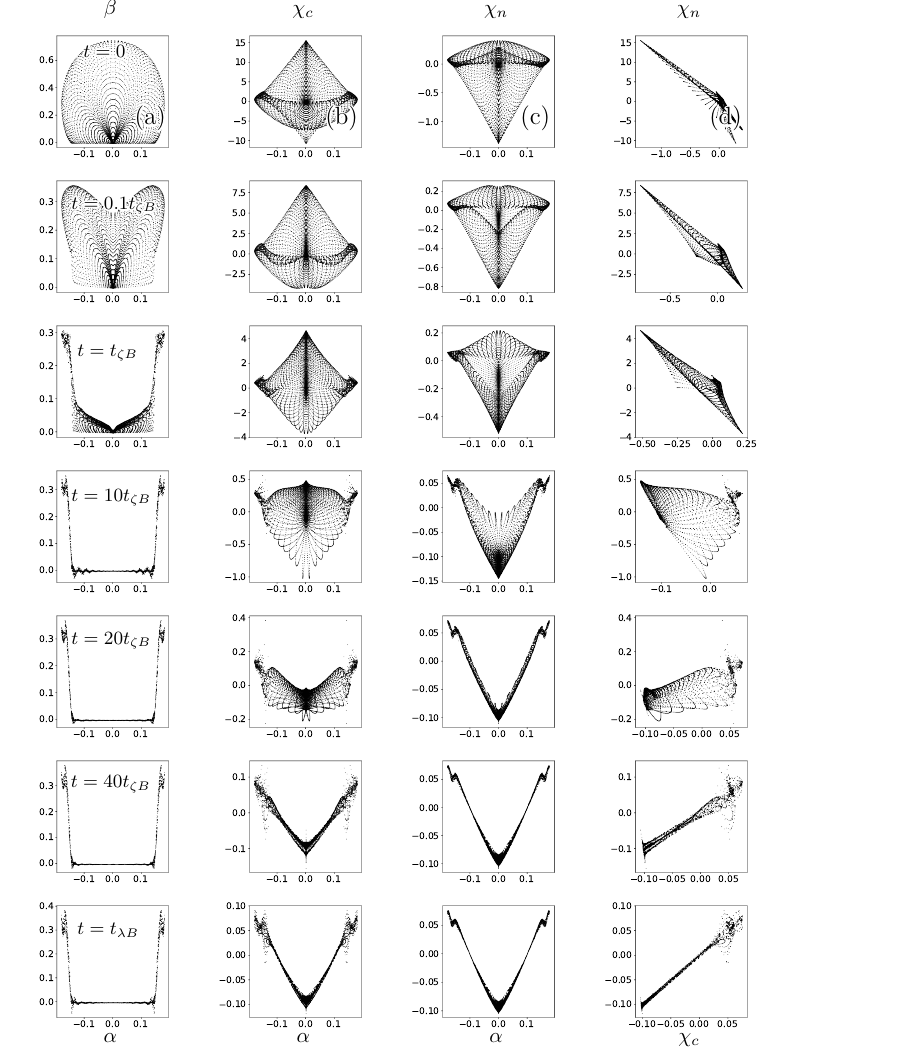}
    \caption{Scatter plot for the same simulation shown in Figs.~\ref{fig4} to \ref{fig6}, with $\zeta = 10^{-3}$ and initial conditions given by Model 2, with the variable on the vertical axis written on top and the variable on the horizontal axis written on the bottom of each column:  
    (a) $\beta$ as function of $\alpha$, (b) $\chi_c$ as function of $\alpha$, (c) $\chi_n$ as function of $\alpha$, (d) $\chi_n$ as function of $\chi_c$.
    We show the relations at all the grid points at $t= 0$, $0.1t_{\zeta B}$, $t_{\zeta B}$, $10t_{\zeta B}$, $20 t_{\zeta B}$, $40t_{\zeta B}$ and $ t_{\lambda B}$.} 
    \label{fig7}
\end{figure*}

Figures~\ref{fig5}(b,c) and \ref{fig6}(b) show that, within a few times $t_{\zeta B}$, a hydromagnetic quasi-equilibrium is reached, in which the fictitious force becomes negligible and all the real forces are close to balancing each other. This implies that the toroidal magnetic force vanishes, $\vec{f_{B}^{\text{Tor}}}\approx 0,$ because there is no toroidal pressure gradient or gravitational force available to balance it. Thus, from equation~(\ref{eqfBTor}), $\Grad \alpha \parallel \Grad \beta$, so one potential is (at least locally) a function of the other, $\beta = \beta (\alpha)$. 
One can see this in the first panel from left to right of Fig.~\ref{fig4}, where the color scale, indicating the values of $\beta$, come progressively closer to matching the field lines, which correspond to surfaces of constant $\alpha$. Perhaps even more clearly, Fig.~\ref{fig7} shows that at early stages there is no clear relation between $\alpha$ and $\beta$, but at $t\gtrsim t_{\zeta B}$ there is an evident dependence. Once this happens, equation~(\ref{eqfBPol}) can be rewritten as 
\begin{equation}\label{eqfBPol1}
    \vec{f}_{B}^{\,\text{Pol}}=-\frac{\Delta^*\alpha+\beta\beta'}{4\pi r^2\sin^2\theta}\Grad\alpha.
\end{equation}
(Here and below, primes denote derivatives with respect to $\alpha$.)
At later times ($t\gtrsim 20\, t_{\zeta B}$), $\chi_{n}$ also becomes roughly a function of $\alpha$, because Urca reactions have reduced the chemical potential perturbations of the charged particles, $\chi_c$, so the poloidal Lorentz force is mostly balanced by the neutrons, $\vec f_{B}^{\text{Pol}}\approx -\vec f_n=n_n\mu\Grad\chi_n$, as seen in the last two rows of Fig.~\ref{fig5}. Fig.~\ref{fig7} also shows that, at $t\sim t_{\lambda B}$, there is a clear relation $\chi_{c}\approx \chi_{c}(\alpha)$, and the NS core is very close to chemical equilibrium, $\chi_{c}\approx \chi_{n}$, so the matter in the core behaves as a barotropic fluid. Therefore, taking into account that, first, $\beta$ becomes a function of $\alpha$, and later, $\chi_n=\chi_c\equiv\chi(\alpha)$, the poloidal force balance equation, equation~(\ref{eqForcebalance}), reduces to a ``Grad-Shafranov (GS) equation'' \citep{gradrubin54,shafranov66}:
\begin{equation}\label{eqGS}
    \Delta^{*}\alpha+\beta \beta' + 4 \pi r^{2} \sin^{2}\theta \,\mu(r) n_{b}(r)\chi^{\prime} = 0.
\end{equation}
We emphasize that this equation is not satisfied in the previous stages of the evolution, which are also hydromagnetic quasi-equilibrium configurations, but out of chemical equilibrium. In fact, as we shall see in \S~\ref{sec:Results}, this state is never reached in the actual evolution of a NS core in the strong-coupling regime.

\subsection{Magnetic energy dissipation}\label{sec:MagneticDiss}

\begin{figure}
    \centering
    \includegraphics[width=8.3cm]{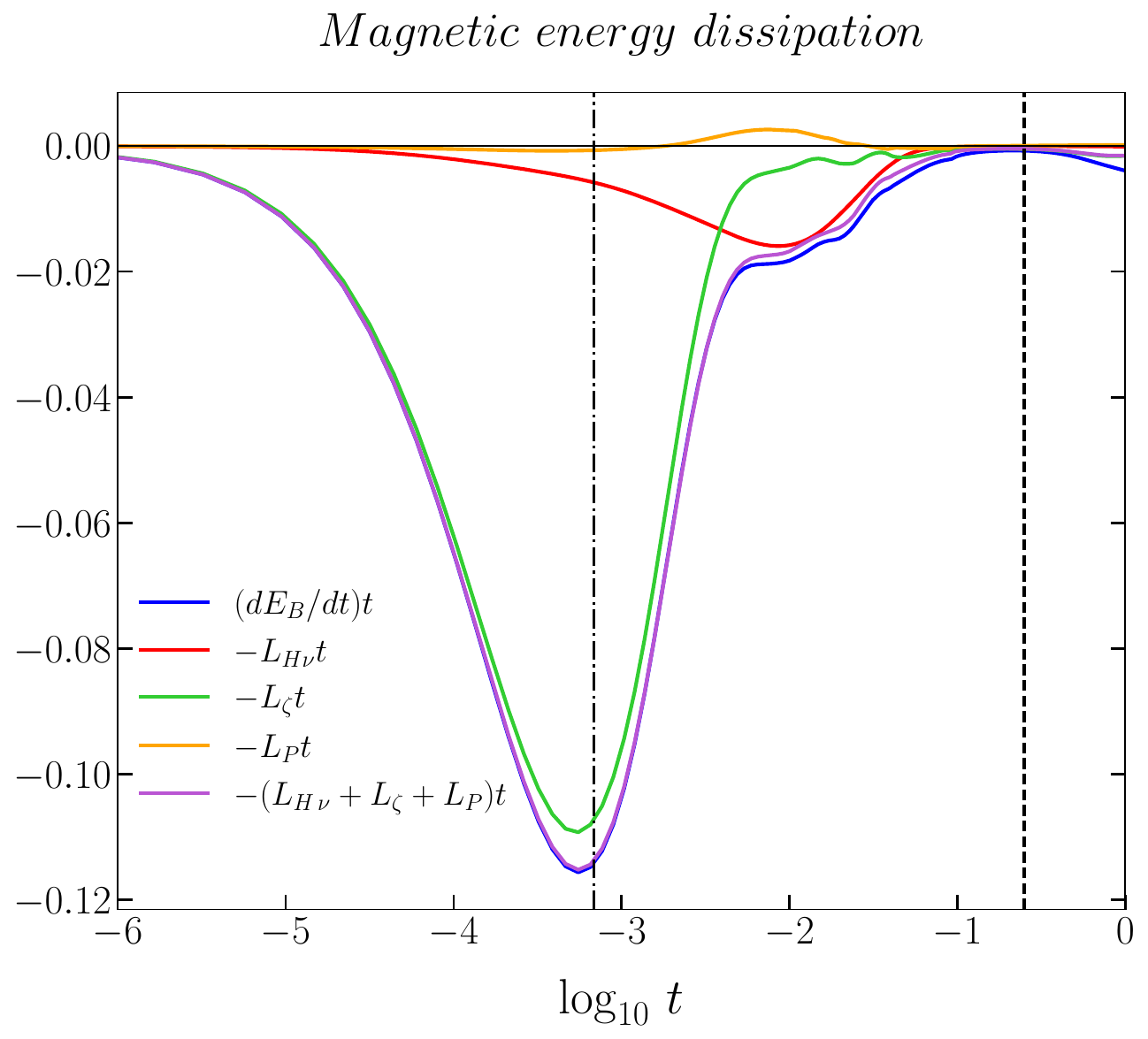}
    \caption{Magnetic energy dissipation for the same simulation as illustrated in the previous figures, whose initial condition is Model 2 and with $\zeta=10^{-3}$, corresponding to $t_{\lambda B}/t_{\zeta B}=370$. The plot shows all the dissipation terms (equation~ \ref{eqEdotB}): the time derivative of the magnetic energy inside the core, $(dE_{B}/dt) t$ (blue); the chemical power released per unit time due to non-equilibrium Urca reactions, $L_{\text{H}\nu}t$ (red); the artificial friction dissipation, $L_{\zeta}t$ (green); the Poynting flux, $L_{P}t$ (orange); and the combination $-(L_{H\nu}+L_{\zeta}+L_{P})t$ (purple), which is used to check energy conservation, as it should be equal to $dE_{B}/dt$. We note that all these variables are multiplied by time, so they represent energy released per unit $\ln t$. The vertical lines correspond, from left to right, to the time scales $t_{\zeta B}$ (dashed-dotted) and $t_{\lambda B}$ (dashed), respectively, while the horizontal solid line corresponds to zero.}
    \label{fig8}
\end{figure}

In this section we discuss the magneto-chemical evolution in terms of the magnetic energy dissipated in the core. We start by deriving the expression for the time-derivative of the magnetic energy contained in the NS core for the most general case, i.~e., considering both the effects of ambipolar diffusion and non-equilibrium Urca reactions, and also allowing a non-linear dependence of $\Delta\Gamma$ on $\Delta\mu/(\pi k_{B}T)$ (see e.~g. \citealt{Reisenegger1995,Fernandez2005,gusakov2005thermal}).

Taking the time derivative of the total magnetic energy contained in the core, 
$E_B = \int_{\cal{V}}B^{2}\,d^{3}x/(8\pi)$,
 where $\cal{V}$ is the core volume; and using equation~(\ref{eqFaraday}), one obtains
\begin{equation}
     \dfrac{d E_B}{dt}  
     =\dfrac{1}{4 \pi}\int_{\cal{V}}\vec{B}\cdot \Grad\times\left(\vec{v}_{c}\times \vec{B}\right)\,d^{3}x.
\end{equation}  
Integrating by parts, 
\begin{equation}
    \dfrac{d E_B}{dt} =-\int_{\cal{V}}\vec{f}_B\cdot\vec{v}_{c}
    \,d^{3}x -\dfrac{1}{4\pi}\oint_{\partial \cal{V}} \vec{B}\times \left(\vec{v}_{c}\times \vec{B}\right)\cdot \vec{dS},
\end{equation}
where $d\mathbf{S}$ is a surface element outward normal to the surface $\partial V$ defined by the crust-core interface. We identify the first term on the right-hand side of this expression as the mechanical work done by the fluid against the Lorentz force, and the second term 
(without the minus sign) as the outward Poynting flux 
integrated over the surface, which we denote hereafter as $L_{P}$. From our equations in \S~\ref{secGeneralEquations}, one obtains
\begin{equation}\label{eqJoule}
\int_{\cal{V}}\vec{f}_B\cdot\vec{v}_{c}
\,d^{3}x=L_{H\nu} +L_{ad} +L_{\zeta},
\end{equation}
where
\begin{gather}
    L_{H\nu}=\int_{\cal{V}}  \Delta \Gamma \Delta \mu\,d^{3}x,\label{eqLHnu}\\
      L_{ad}=\int_{\cal{V}} \gamma_{cn}n_{c}n_{n}|\vec{v}_{\text{ad}}|^{2}\,d^{3}x,\label{eqLa}\\
      L_{\zeta}=\int_{\cal{V}}\zeta n_{n} |\vec v_{n}|^{2}\, d^{3}x\label{eqLzeta}.
\end{gather}
are the power dissipated by non-equilibrium Urca reactions, ambipolar diffusion, and artificial friction, respectively. When deriving  
equation (\ref{eqJoule}), the terms proportional to $\psi/c^{2}$ were disregarded, in constancy with Newtonian gravity.
Nonetheless, we computed them numerically and checked their insignificance when compared to the other terms. 
Therefore, the rate of change of the magnetic energy stored in the core reads as
\begin{equation}\label{eqEdotB}
    \dfrac{dE_B}{dt}=-L_{H\nu}-L_{ad}-L_{\zeta}-L_{P}.
\end{equation}
 
Figure~\ref{fig8} shows all the terms that dissipate energy in our reference simulation for the strong-coupling regime, in which
$L_{ad}=0$ and $\Delta\Gamma=\lambda\Delta\mu$. As might have been expected, initially the fictitious friction dissipation $L_{\zeta}$ dominates, since the dynamics is given by quick toroidal fluid motions with velocity $\vec{v}^{\text{Tor}} =\vec{f_{B}}^{\text{Tor}}/(\zeta n_{n})$, 
which unwind the magnetic field 
around $t\sim t_{\zeta B}$. This occurs at the position of the largest peak of $(dE_B/dt)\, t$, corresponding to where most of the magnetic energy dissipates. At later times ($\gg t_{\zeta B}$), the real physical evolution  
begins, when non-equilibrium Urca reactions become effective and the chemical power released by them, $L_{H\nu}$, is the main source of magnetic energy dissipation in the second, lower peak. A curious fact of our simulations is the 
opposite sign of the Poynting flux $L_{\text{P}}$ with respect to the other dissipative terms, also seen in Fig.~\ref{fig8}, which indicates that external magnetic energy (although a small amount) is being injected into the NS core. One might have anticipated such behavior from the fact that outside the core we are imposing a current-free magnetic field configuration, so the magnetic dissipation occurs only in the core. This is analogous to a wire carrying an electric current: Since most of the magnetic energy is located outside, but the dissipation takes place inside, there is a flux of magnetic energy from outside into the wire.
\begin{figure*}
    \centering
    \includegraphics[width=15cm]{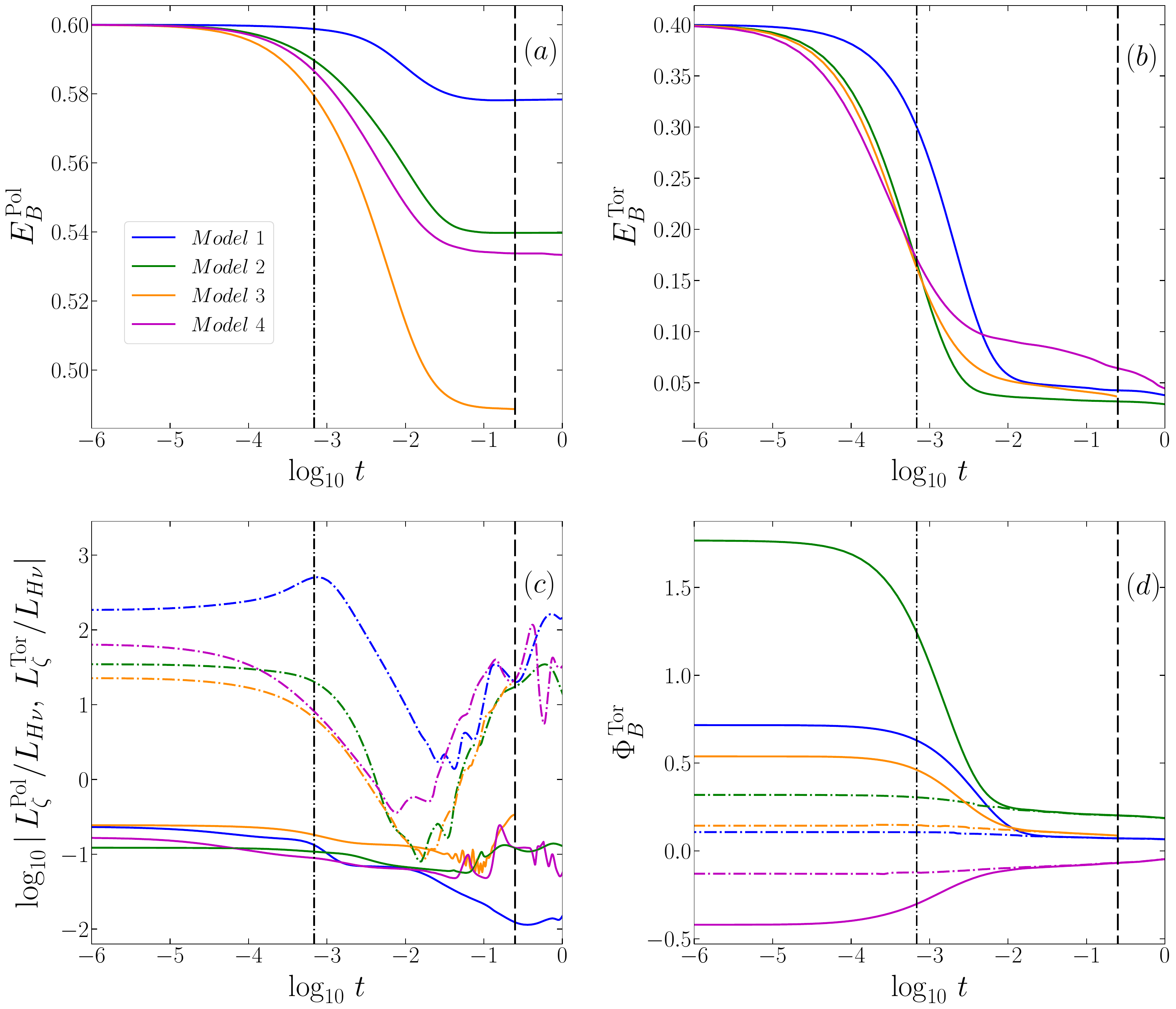}
    \caption{
    Evolution at constant temperature for the different initial magnetic field configurations listed in Table~\ref{table1} (with color coding as shown in panel [a]), always with $\zeta=10^{-3}$, showing:  (a) poloidal and (b) toroidal magnetic energy, both in units of the total initial magnetic energy; (c) quotient between artificial friction dissipation, split into a poloidal and a toroidal parts; the chemical power released due to non-equilibrium Urca reactions, 
    $L_{\zeta}^{\text{Pol}}/L_{\text{H}\nu}$ (solid) and $L_{\zeta}^{\text{Tor}}/L_{\text{H}\nu}$ (dashed); and (d) toroidal magnetic fluxes, where the solid lines are the total fluxes and the dashed-dotted are the fluxes computed only in the regions of closed poloidal field lines. 
    The vertical lines correspond, from left to right, to the time scale $t_{\zeta B}$ (dashed-dotted), and $t_{\lambda B}$ (dashed), respectively.}
    \label{fig9}
\end{figure*}

Our simulations also allow us to determine how much magnetic energy is dissipated in order to establish the final GS equilibrium for different initial conditions. Fig.~\ref{fig9}(a,b) shows the evolution of the internal poloidal and toroidal magnetic energies for the different models we simulated. These models are specified in Table~\ref{table1} and Fig.~\ref{fig2}).

The results show that during the early evolution, $t\lesssim t_{\zeta B}$, most of the toroidal energy is dissipated by almost the same amount for all the magnetic field models. This strong dissipation is due to toroidal fluid motions generated by the magnetic field in order to reach the hydromagnetic quasi-equilibrium and thus satisfy the condition $\beta = \beta (\alpha)$. As these toroidal fluid motions  proceed, they are opposed by the artificial friction force, so that most of the dissipation is caused initially by the artificial friction, $L_{\zeta}^{\text{Tor}}\gg L_{\text{H}\nu}$ (Fig.~\ref{fig9}[c]), where $L_{\zeta}^{\text{Tor}}$ is computed considering only the toroidal component of the velocity (see equation~\ref{eqLzeta}).

We remark that this pronounced initial evolution of the toroidal energy is a consequence of starting with an initially uncompensated toroidal magnetic field. If the simulations were to begin with a quasi-equilibrium field configuration ($\beta = \beta (\alpha)$), the toroidal field would not undergo significant changes. This can be corroborated in Fig.~\ref{fig9}(d), where we computed the toroidal magnetic flux

\begin{equation}
 \Phi_{B}^{\text{Tor}}=\int_{ \cal{S}} \vec{B}^{\text{Tor}}\cdot \vec{dS},
\end{equation}
considering two choices for the surface ($\cal{S}$) of integration: the entire meridional cross-section of the NS core, and the piece corresponding to the closed poloidal field lines. 

The results show that, as expected, the evolution of the total toroidal flux is particularly significant around $t\sim t_{\zeta B}$ as the core approaches hydromagnetic quasi-equilibrium. Conversely, the toroidal flux associated with the regions of closed poloidal field lines appears to remain nearly conserved throughout the entire evolution. Therefore, if the initial magnetic field were to be close to a hydromagnetic quasi-equilibrium state, the toroidal field would indeed undergo minimal evolution in order to reach the final GS equilibrium state. All in all, by far the main effect of the artificial friction is to eliminate the toroidal magnetic field in the regions of open poloidal field lines.
On the other hand, during this early evolution, the poloidal energy evolves as it is more sensitive to Urca reactions, which operate on the time scale $\sim t_{\lambda B}$.

During the late evolution, $t_{\zeta B}\lesssim t \lesssim t_{\lambda B}$,
the toroidal magnetic energy evolves very little and settles to a common asymptotic state for all the magnetic field models. 
For the poloidal energy, however, most of the dissipation occurs in this late evolution, when Urca reactions operate, as can be seen in Fig.~\ref{fig9}(a) and, more clearly, for the simulation with $\zeta =10^{-4}$ in Fig.~\ref{fig3}(a,b).
One can see also that, for all initial configurations, only a small fraction of the poloidal energy is dissipated. This is particularly evident in Model 1, which corresponds to the simplest dipole model. 
On the opposite extreme, Model 3 exhibits the most pronounced decrease in poloidal energy, likely due to its less symmetric nature.

In terms of the dissipation, at this late stage of the evolution, the artificial friction should be irrelevant, i.~e., $L_{\text{H}\nu}\gg L_{\zeta}^{\text{Pol}}$ and $L_{\text{H}\nu}\gg L_{\zeta}^{\text{Tor}}$, as the system has already reached the hydrostatic equilibrium, and since $L_{\zeta}^{\text{Tor}}\propto (\Grad \beta \times \Grad \alpha)^{2}$. Unfortunately, numerical noise affects the last moments of our simulations $\gtrsim t_{\lambda B}$, and $L_{\zeta}^{\text{Tor}}$ starts to be dominant again (see Fig.~\ref{fig9} [c]), however at this point the core is very close to chemical equilibrium and all dissipation effects become very small.

Therefore, we conclude this section by noticing that our results suggest that, for relatively smooth initial magnetic field configurations in axial symmetry, reaching the GS equilibrium requires dissipating only a few percent of the poloidal and and most of the toroidal magnetic energies.

\section{Magneto-thermal evolution}\label{sec:Tev}

Up to this point, we have solved equations (\ref{eqIndA})-(\ref{eqDmu}) in the unphysical situation where the core temperature remains constant, i.~e., when $\lambda$ does not change in time. However, in reality, the core temperature evolves, and since $\lambda$ depends strongly on the temperature, it will also evolve. Below, we describe an efficient strategy to include the thermal evolution, using the output of the simulations at constant temperature and incorporating the thermal evolution by reparameterizing the time variable.

\subsection{Temperature evolution}\label{sec:TemperatureEvolution}

Before describing the procedure, we introduce the equation needed to evolve the core temperature, which is given by a thermal balance equation in the isothermal approximation \citep{Thorne1977,yakovlev2004neutron}
\begin{equation}\label{eqDTdt}
    \dfrac{dT}{dt}=\dfrac{1}{C}\left(L_{H}-L_{\nu}-L_\gamma \right),
\end{equation}
where $L_{H}$ is the total power released by heating mechanisms, $L_{\nu}$ is the total neutrino luminosity, $L_\gamma$ is the luminosity associated to the photons emitted from the stellar surface, and $C$ is the total heat capacity of the NS core, which for our stellar model is given by 
\begin{equation}
    C = 1.6\times 10^{39}T_9 
    \,\text{erg}\,\text{K}^{-1},
\end{equation}
where $T_9=T/(10^9\text{K})$.

In the present case, equation~(\ref{eqDTdt}) takes the form
\begin{equation}\label{eqdTdtStrongcoupling}
    \dfrac{dT}{dt} = \dfrac{1}{C}\left(L_{\zeta}+L_{H\nu}- L_{\nu}\right),
\end{equation}
where the only heating mechanisms are the chemical power released by non-equilibrium Urca processes, $L_{H\nu}$; and the artificial friction dissipation, $L_{\zeta}$ (see equation~\ref{eqEdotB}), which was included for consistency. 
We neglected the photon luminosity because this cooling mechanism is only relevant for later stages in the NS life \citep{yakovlev2004neutron}. 

The expressions for the neutrino luminosity and the chemical power released by non-equilibrium Urca reactions are 
\citep{Haensel1992} 
 \begin{gather}             
L_{\nu}=
\int_{\cal{V}} \epsilon_{\nu}^{*}(T)\,F(\xi)\,d^{3}x, 
\label{eqLnu}\\
L_{H\nu}=\int_{\cal{V}}\Delta\Gamma\Delta\mu\,d^{3}x=
\int_{\cal{V}}\epsilon_{\nu}^{*}(T)\, \xi H(\xi)\,d^{3}x,\label{eqDgamma}
 \end{gather}
where $\epsilon_{\nu}^{*}$ is the equilibrium neutrino emissivity, which is a function of the stellar model and temperature, and $F(\xi)$ and $H(\xi)$ are dimensionless control functions of $\xi=\Delta\mu/(\pi k_{B}T)$. The exact expressions for these functions were found by \citet{Reisenegger1995} for both direct and modified Urca processes. Consistently with our stellar model, we assume only modified Urca process, for which these functions for arbitrary values of $\xi$ are given by
\begin{gather}
 F(\xi)= 1 + \dfrac{22020\xi^{2}+5670\xi^{4}+420\xi^{6}+9\xi^{8}}{11513},\label{eq:pm64}\\
H(\xi)=  \dfrac{14680\xi+7560\xi^{3}+840\xi^{5}+24\xi^{7}}{11513}\label{eq:pm65}.
\end{gather}
Since in the present problem we will mostly have $|\xi|\ll 1$ (the ``sub-thermal'' regime) as already assumed in equation~(\ref{eqLambda}), the net heating rate becomes 
\begin{equation}\label{eqLnuapprox}
    L_{H\nu}-L_\nu=-\int_{\cal{V}}\epsilon^{*}_{\nu}\left[1+\dfrac{7340}{11513}\xi^2+{\cal O}(\xi^4)\right]\,d^{3}x\approx-L^{*}_{\nu}-\dfrac{1}{2}L_{H\nu},
\end{equation}
where $L^{*}_\nu$ is the equilibrium neutrino luminosity, which for our stellar model reads
\begin{equation}
    L^{*}_\nu = 3.8\times 10^{40}T_9^8 \,\text{erg}\,\text{s}^{-1}.
\end{equation}
Equation~(\ref{eqLnuapprox}) shows that non-equilibrium Urca reactions cause enhanced cooling as a consequence of the magnetic feedback (see also \citealt{Fernandez2005}). However, we will show that no significant magnetic feedback takes place even for very strong magnetic fields.

Incorporating equation (\ref{eqdTdtStrongcoupling}) directly into the existing system of equations (\ref{eqIndA})-(\ref{eqDmu}) for our simulations is the most natural
approach in principle. However, for enhanced efficiency, we employ an alternative strategy outlined in \S~\ref{sec:timerepar}.


\subsection{Time reparametrization}\label{sec:timerepar}

Our previous simulation results, calculated at constant temperature, can be translated into a more realistic evolution with variable temperature, and thus time-dependent $\lambda(r,t)$, by noting that the system of equations (\ref{eqIndA})-(\ref{eqDmu}) remains invariant under the following changes of variables
\begin{gather}
    dt = 
    \frac{\lambda'}{\lambda(t)}dt',\label{rescale-time}\\
    \vec v = 
    \frac{\lambda(t)}{\lambda'}\vec v', \label{rescale-velocity}\\
    \zeta(t) =
    \frac{\lambda'}{\lambda(t)}\zeta', \label{rescale-zeta}
\end{gather}
where primed variables ($\lambda',t',\vec v',\zeta',T'$) correspond to the previous simulations at an arbitrary constant temperature $T'$, while unprimed variables ($\lambda,t,\vec v,\zeta, T$) take into account the thermal evolution and its effect on the parameters. (As seen in equation~\ref{eqLambdaMU}, the functions $\lambda$ and $\lambda'$ have the same dependence on density, and thus on $r$, which thus cancels out in their ratio. Therefore, we wrote the latter simply as $\lambda'/\lambda(t)$, which depends only on temperature, and hence on time.)  

Thus, it is possible to solve the system of equations~(\ref{eqIndA})-(\ref{eqDmu}) for time-independent
$\lambda'(r)$ and $\zeta'$, then solve for the evolution of the temperature in order to obtain the time dependence of $\lambda(r,t)$, and finally reparametrize the time variable in order to obtain the evolution of all variables for the realistic case of variable temperature. Our self-consistent approach can be understood as follows:

\begin{enumerate}
    \item We run the simulations that evolve the set of equations (\ref{eqIndA})-(\ref{eqDmu}) at constant temperature (thus time-independent $\lambda'$ and $\zeta'$) as presented in \S~\ref{secModel}, calling the time variable $t'$.
    \item Then, using these results, we solve the equation for the temperature as a function of $t^{\prime}$, which with the aid of equation~(\ref{eqLnuapprox}) reads as
        \begin{equation}\label{e71}
            \dfrac{dT}{dt'} = \dfrac{1}{C(T)}\left[L_{\zeta}^{\prime}-\dfrac{1}{2}L_{H\nu}^{\prime}- L^{*\prime}_{\nu}\left(\dfrac{T}{T'}\right)^{2}\right],
    \end{equation}
  where $L^{*\prime}_{\nu}$ is the equilibrium luminosity evaluated at the constant reference temperature $T'$. Here, we also used $\lambda/\lambda^{\prime}=(T/T')^{6}$, and the scaling in equations~(\ref{rescale-time})-(\ref{rescale-zeta}), so that $L_{H\nu} = L_{H\nu}^{\prime}(T/T')^{6}$, $L_{\zeta} = L_{\zeta}^{\prime}(T/T')^{6}$, and $L^{*}_{\nu}=L^{*\prime}_{\nu}(T/T')^{8}$ where the primed variables are evaluated at the fixed temperature $T'$, not at the true, time-dependent temperature $T$.
    
    \item Finally, we obtain the physical time variable that includes the effects of the temperature evolution by integrating equation~(\ref{rescale-time}). Thus, we can plot any variable of interest as a function of $t$, the real physical time.
\end{enumerate}

\subsection{Analytic solution for passive cooling}\label{sec:passive cooling}

If there is no substantial feedback (heating or enhanced cooling) due to the magnetic field evolution, the thermal evolution equation 
(\ref{eqdTdtStrongcoupling}) reduces to $dT/dt=-L_\nu^*(T)/C(T)$, or equivalently $dT_9/dt=-T_9^7/(1.32\text{yr})$, so the temperature evolves as
\begin{equation}\label{eqTpassive1}
T(t) = \dfrac{T_{init}}{(1+t/t_{cool})^{1/6}},
\end{equation}
where $T_{init}$ is the initial temperature, and
\begin{equation}    
t_{cool}=0.22\left(\dfrac{T_{init}}{10^{9}\,\text{K}}\right)^{-6}\,\text{yr},
\end{equation}
is a characteristic cooling time at $T=T_{init}$.

For the initial temperature used in our simulations, $T_{init}=2\times 10^{10}\,\text{K}$, $t_{cool} =3.4\times 10^{-9}\,\text{yr}$, and, for any realistically large initial temperature, $T_{init}\gg T_{trans}\approx 5\times 10^8\,\mathrm K$, the transition from the strong-coupling to the weak-coupling regime occurs at $t_{trans}\sim 30\,\mathrm{yr}$.

\begin{figure}
    \centering
    \includegraphics[width=8.6cm]{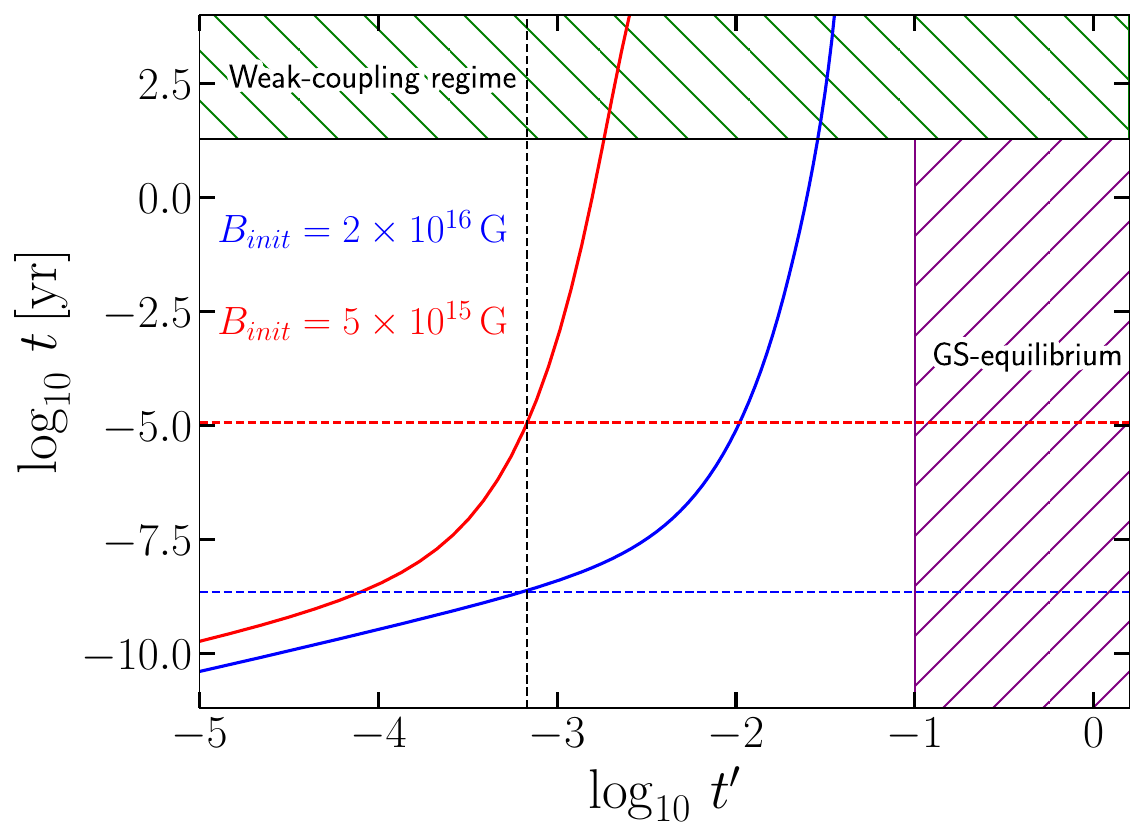}
\caption{Physical time variable $t$ (in years) for evolving temperature as a function of auxiliary time $t'$ (in code units) for constant temperature, as given by equation~(\ref{eqTTprime}), for two different initial rms magnetic field strengths: $B_{init}=2\times 10^{16}\,\mathrm{G}$ (blue) and $B_{init}=5\times 10^{15}\,\mathrm{G}$ (red). The dashed vertical line marks the time scale $t'_{\zeta B}$, which is independent of $B_{init}$. The horizontal lines show, from bottom to top, the time scales $t_{\zeta B}= t(t'_{\zeta B})$ (dashed) mapped from equation~(\ref{eqttprime2}) for the different initial rms magnetic field strengths with their respective colors, and the transition time $t_{trans}$ from the strong-coupling to the weak-coupling regime (black-solid). 
The purple-hatched area displays the time in which GS-equilibrium is reached in our simulations at constant temperature, and the green-hatched area shows the time range corresponding to the weak-coupling regime. 
}
\label{fig10}
\end{figure}

Applying the time rescaling of equation~(\ref{rescale-time}) with the temperature given by equation~(\ref{eqTpassive1}), the physical time variable $t$ for the evolution with variable temperature becomes an exponential function of the auxiliary time $t'$ used for the constant-temperature evolution, 
\begin{equation}\label{eqTTprime}
t=t_{cool}\left\{\exp\left[\frac{t^{\prime}}{0.22\,\mathrm{yr}}\left(\frac{T'}{10^9\,\mathrm{K}}\right)^6\right]-1\right\}=t_{cool}\left[\exp\left(3\,\frac{t^{\prime}}{t'_\lambda}\right)-1\right].
\end{equation}
(Here and below, all time scales with primes refer to times in our constant-temperature simulations of \S~\ref{results_contantT}.) Writing $t'$ in terms of the code units defined in Table~\ref{table2}, this reduces to
\begin{equation}\label{eqttprime2}
t=t_{cool} \left\{\exp{\left[3.0\times 10^5\left(\frac{B_{init}}{10^{15}}\right)^{-2}t'\right]}-1\right\},
\end{equation}
which is plotted in Fig.~\ref{fig10} for two large values of $B_{init}$. 
Clearly, moderate values of $t'\,(\lesssim 1)$ and $B_{init}\,(\lesssim 10^{15}\,\mathrm{G})$ yield huge values of $t$, many orders of magnitude larger than $t_{trans}$ (or even than the age of the Universe), indicating that very little evolution will occur during the strong-coupling regime, $t\lesssim t_{trans}$. Only for much larger field strengths, a substantial evolution might occur.

As discussed in \S\S~\ref{sec:timescales} and \ref{sec:artificialfriction}, the initial stages $t'\lesssim t'_{\zeta B}$ of our constant-temperature simulations are dominated by the fictitious friction force, which allows to set up a hydromagnetic equilibrium that in a real NS core would be reached in a few Alfv\'en times. Therefore, only for later times, $t'\gg t'_{\zeta B}$, the simulations can be expected to represent the true astrophysical evolution. For $t'_{\zeta B}$ to correspond to a physical time earlier than the transition to the weak-coupling regime, we must have
\begin{equation}\label{eq:Bfortzetalessthanttrans}
    B_{init}\gtrsim 6\times 10^{15}\,\mathrm{G}\left[\frac{\zeta'/10^{-3}}{\ln(T_{init}/T_{trans})}\right]^{1/2}\sim 3\times 10^{15}\left(\frac{\zeta'}{10^{-3}}\right)^{1/2}\,\mathrm{G},
\end{equation}
where $\zeta'=10^{-3}$ was the dimensionless friction coefficient used in most of our simulations. Thus, our simulations represent the astrophysically relevant regime only for very strong magnetic fields satisfying these conditions. Figure~\ref{fig10} illustrates the correspondence between the time scale $t'_{\zeta B}$ and real time variables $t_{\zeta B}$, highlighting the increasing magnitude of the latter values as we diminish the initial rms magnetic field strength. 

On the other hand, we have seen in our constant-temperature simulations (Figs.~\ref{fig3}, \ref{fig6}, \ref{fig7}, \ref{fig8}, \ref{fig9}) that the time needed to reach the final chemical and GS equilibrium typically falls in the range $t_{GS}'\approx 0.1 - 1$ in code units, depending on the magnetic field model. Thus, the strength of the initial magnetic field required to reach this equilibrium before the transition to the weak-coupling regime is given by
\begin{align}
&B_{init}\gtrsim 2\times 10^{17}\,\mathrm{G}\left[\frac{t_{GS}'}{\ln(T_{init}/T_{trans})}\right]^{1/2} \sim 10^{17}\,{t_{GS}'}^{1/2}\,\mathrm{G},&
\end{align}
which is unrealistically large even for magnetars (see Fig.~\ref{fig10}). Thus, we expect that no substantial evolution will occur in the strong-coupling regime, unless there is strong feedback that might change the temperature evolution. We verify below that no such feedback occurs for realistic magnetic fields.

\begin{figure}
    \centering
    \includegraphics[width=8.2cm]{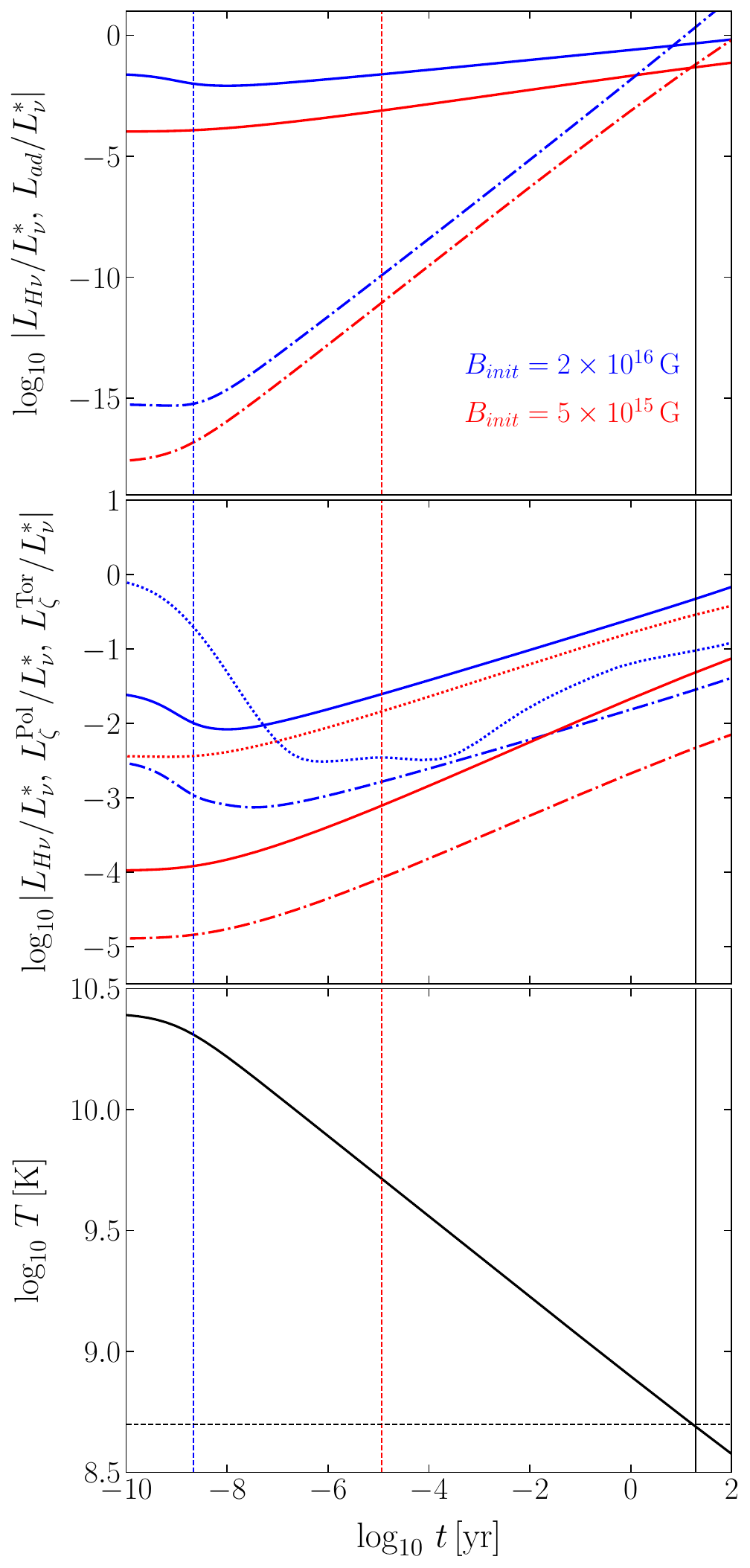}
\caption{Magneto-thermal evolution of the relevant variables for the same simulation presented in \S~\ref{sec:magchem} (Model 2 with $\zeta^{\prime}=10^{-3}$) after applying the time reparametrization procedure for two different rms initial magnetic field strengths ($B_{init}=2\times 10^{16}\,\mathrm{G}$ in blue, $B_{init}=5\times 10^{15}\,\mathrm{G}$ in red), and with the initial temperature $T_{init} =2\times 10^{10}\,\text{K}$. Upper panel: the chemical power released by non-equilibrium Urca reactions, $L_{H\nu}$ (solid), and ambipolar dissipation, $L_{ad}$ (dashed-dotted), both normalized to the equilibrium luminosity $L_{\nu}^{*}$. Middle panel: the chemical power released, $L_{H\nu}$ (solid), and the dissipation due to artificial friction split into toroidal, $L_{\zeta}^{\text{Tor}}$ (dotted), and poloidal, $L_{\zeta}^{\text{Pol}}$ (dashed-dotted), contributions, normalized to the equilibrium luminosity $L_{\nu}^{*}$. Lower panel: The thermal evolution for the case of passive cooling, i.~e., solving $dT/dt=-L_\nu^*/C$. (The evolution curves for the two magnetic field strengths are indistinguishable from the passive cooling curve.) The vertical lines show, from left to right, the time scales $t_{\zeta B}$ mapped from equation~(\ref{eqTTprime}) as $t_{\zeta B} = t(t'_{\zeta B})$ for the two values of $B_{init}$ (dashed and with the respective colors), and the transition time (solid black) when the core reaches the transition temperature $T_{trans} = 5\times 10^{8}\,\text{K}$, while the dashed horizontal line in the lower panel shows $T_{trans}$.
}
\label{fig11}
\end{figure}

\subsection{Simulation results}\label{sec:Results}

\begin{figure*}
    \centering
    \includegraphics[width=17.8cm]{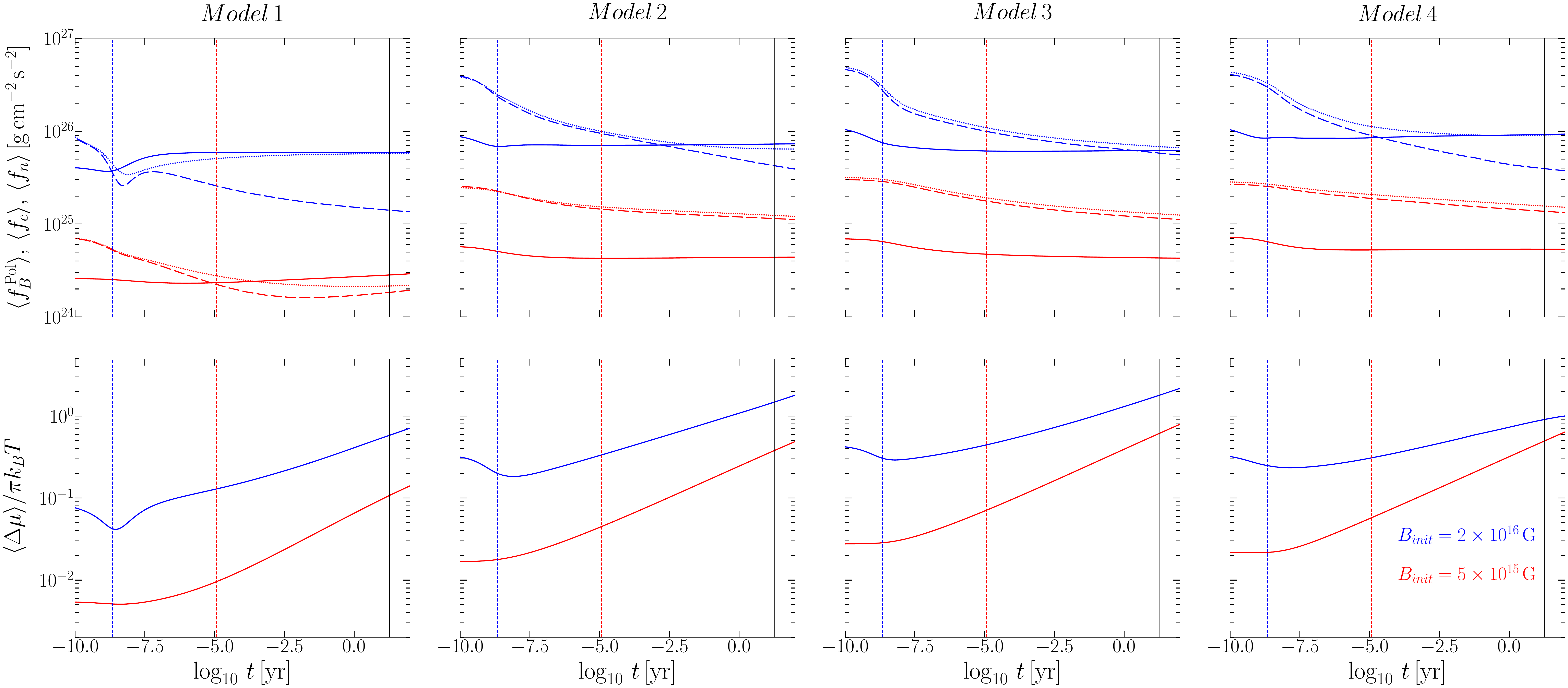}
\caption{Evolution of the rms-average of the forces (top),$\langle f_{B}^{\text{Pol}} \rangle$ (solid), $\langle f_{c}\rangle$ (dashed) and $\langle f_{n} \rangle$ (dotted), and $\langle \Delta \mu \rangle/\pi k_{B}T$ (bottom) for the different magnetic field strengths and models with $T_{init}=2\times10 ^{10} \,\text{K}$, and $\zeta^{\prime}=10^{-3}$. The vertical lines shows, from left to right, the time scale $t_{\zeta B}$ (dashed and with the respective colors) mapped from equation~(\ref{eqTTprime}) as $t_{\zeta B} = t(t'_{\zeta B})$, and the transition time $t_{trans}$ (black-solid) at which the core reaches the transition temperature $T_{trans} = 5\times 10^{8}\,\text{K}$. }
\label{fig12}
\end{figure*}

In this subsection, we analyze the evolution of the progressively cooling NS core in the strong-coupling regime by applying the time reparametrization $t(t')$ described in \S~\ref{sec:timerepar} to the simulations run at constant temperature and discussed in \S~\ref{results_contantT}. These can represent the true astrophysical evolution only for $t'\gg t'_{\zeta B}$, when they are no longer dominated by artificial friction, and $t\lesssim t_{trans}$, before the true evolution would become dominated by ambipolar diffusion. In order to leave a non-negligible time interval in which both conditions are satisfied, we choose extremely strong initial magnetic fields compatible with equation~(\ref{eq:Bfortzetalessthanttrans}), noting that for much weaker fields such as found even in magnetars the evolution will be slowed down exponentially.

Figure~\ref{fig11} (upper and middle panels) shows the evolution of all the relevant luminosities normalized with respect to the equilibrium neutrino luminosity, $L_{\nu}^{*}$, as functions of true physical time $t$, after applying the time reparametrization procedure for model 2.
Although we do not consider the effect of the ambipolar velocity on the evolution of the magnetic field, we compute the ambipolar heating $L_{ad}$, equation~(\ref{eqLa}), with $\vec{v}_{ad}$ given by equation~(\ref{eqVad}), in order to estimate how strong the magneto-thermal feedback becomes as the weak-coupling regime is approached around $t=t_{trans}$. Clearly, $L_{H\nu}$, $L_{ad}$, and $L_{\zeta}$ are smaller than $L_{\nu}^{*}$ almost throughout the whole strong-coupling regime, for both of the very high magnetic field strengths shown. As a consequence, there is no substantial magnetic feedback on the thermal evolution, justifying the assumption of passive cooling made in \S~\ref{sec:passive cooling}. In fact, the temperature evolution for the case of passive cooling, shown in the lower panel of Fig.~\ref{fig11}, is indistinguishable from the curves obtained self-consistently with feedback from the two magnetic field strengths considered in the other panels. This can also be understood from the lower row of Fig.~\ref{fig12}, where $\langle \Delta \mu \rangle/\pi k_B T < 1$, except for the higher magnetic field strength $B_{init}=2\times10^{16}\,\text{G}$ and the latest stages of the strong-coupling regime for Models 2 and 3. The feedback is expected to become much weaker still for weaker, more realistic magnetic field strengths.

Therefore, throughout the strong-coupling regime, the thermal evolution is nearly unaffected by that of the magnetic field, while the latter is strongly affected by the former through the net reaction rate, $\Delta\Gamma=\lambda\Delta \mu$, with $\lambda \propto T^{6}$. 
As the NS core passively cools, Urca reactions become progressively less effective in restoring chemical equilibrium, and the whole magneto-chemical evolution is slowed down. 
As discussed in \S\S~\ref{sec:logtev} and \ref{sec:passive cooling}, the evolution becomes slower for weaker magnetic fields, as supported by
Fig.~\ref{fig12}, where only the stronger
magnetic field generates a pronounced evolution of the forces. 

We also note that, in the upper row of Fig.~\ref{fig12}, the moment when the three forces are of the same order, $|\vec{f}_n|\sim|\vec{f}_c|\sim|\vec{f}^{\text{Pol}}_{B}|$, occurs at very different times for the four models we considered. We checked, however, that these differences are only a factor $\sim 2-3$ in the constant-temperature time variable $t^{\prime}$, whereas the exponential dependence of $t$ on $t^{\prime}$ in equation (\ref{eqTTprime}) amplifies these time differences in the time variable $t$, making the magneto-thermal evolution for very strong $B_{init}$ quite sensitive to the magnetic field geometry.


\section{Conclusions}\label{sec:conclusion}

In this paper, we have reported the first simulations of the magneto-thermal evolution of a NS core in the strong-coupling regime applicable at high temperatures, $T>T_{trans}\approx 5\times 10^8\,\mathrm{K}$, covering two relevant aspects not considered before, namely the interplay between non-equilibrium Urca reactions and magnetic field dynamics as well as the effect of the thermal evolution on the evolution of the magnetic field. The latter was included by starting from the results of the simulations at constant temperature and implementing an efficient numerical strategy to translate these results to the realistic case of evolving temperature. Like previous work \citep{Castillo2017,castillo2020twofluid}, 
these simulations are done in axial symmetry, for a two-fluid (neutrons and charged particles) NS core surrounded by a non-conducting medium, and include a fictitious friction force that allows the system to quickly reach a hydromagnetic equilibrium state while ignoring the inertial terms and, in this case, the time derivatives in the particle conservation laws, that would substantially slow down the simulations. The main conclusions can be summarized as follows:
\begin{enumerate}
  \item In the initial stages, in which the fictitious friction is important, the NS core evolves mostly through toroidal motions that unwind the magnetic field lines, leading to a hydromagnetic equilibrium state in which the toroidal component of the Lorentz force vanishes, confining the toroidal magnetic field component to the region of closed poloidal field lines. In a real NS core, this process is expected to happen very quickly, within a few Alfv\'en times.

  \item \emph{In the constant-temperature simulations}, the magneto-chemical evolution generated by the interplay between Urca reactions and the magnetic field dynamics leads the NS core to evolve towards a chemical equilibrium state in which the fluid becomes barotropic and the magnetic field satisfies the non-linear Grad-Shafranov equation.  
  This state is reached in a time
    \begin{equation}
    t_{\lambda B} \sim (0.6\,-\,6) \times 10^{4}\,\left(\dfrac{B}{10^{15}\text{G}}\right)^{-2} \left(\dfrac{T}{10^{9}\text{K}}\right)^{-6}\,\text{yr},
    \end{equation}
    which for realistic magnetic fields is much larger than the previous, traditionally underestimated
    value $t_\lambda$ provided by equation~(\ref{eqtlambda}) \citep{DuncanThompson1996,lander221protoNS}. Moreover, our results indicate that, once the fluid has settled into a hydromagnetic quasi-equilibrium with vanishing toroidal Lorentz force,
    the magnetic field needs to evolve very little in order to reach the final GS equilibrium state. This may appear surprising at first glance, as one would intuitively expect a significant magnetic evolution accompanied by a strong dissipation during the transition from a wide range of possible equilibria, where two fluid forces proportional to the gradients of two independent scalar functions, $\vec{f}_n=-n_n\mu\nabla\chi_n$ and $\vec{f}_c=-n_c\mu\nabla\chi_c$, balance the poloidal Lorentz force, to the GS equilibrium state, where the latter force is balanced by a force proportional to the gradient of a single scalar function, $\vec{f}_n+\vec{f}_c=-n_b\mu\nabla\chi$. Possibly, the results might be different if we considered more complex magnetic field structures or relaxed the axial-symmetry constraint. 
    
    \item Our results also revealed an intriguing behavior related to the fluid forces: for an arbitrary initial magnetic field, the fluid forces (pressure plus gravity) exerted by the charged particles and the neutrons were found to approximately balance each other, $\vec{f}_{c}\approx-\vec{f}_n$, and were considerably stronger than the poloidal Lorentz force $\vec f_B^{\text{Pol}}$. This observation may seem unexpected since the chemical potential perturbations responsible for the fluid forces originate from $\vec f_B^{\text{Pol}}$. However, this phenomenon arises as an indirect consequence of stable stratification. Specifically, it stems from the fact that the radial density profiles, $n_c(r)$ and $n_b(r)$, although not identical, exhibit similar smooth decreases over a characteristic length scale $\ell_c\sim R_{core}$, substantially larger than the magnetic length scale $\ell_B$.

    \item \emph{When the temperature evolution was included}, our results showed that even the strongest magnetic fields we considered ($B\gtrsim 10^{16}\,\text{G}$) cannot generate an appreciable magnetic feedback on the thermal evolution. Consequently, the NS core cools passively, so the effectiveness of Urca reactions gradually declines, the overall magneto-chemical evolution decelerates, and there is no significant magnetic evolution before ambipolar diffusion becomes more important than Urca reactions, thus ending the strong-coupling regime. This was quantified by showing that the physical time $t$ (corresponding to evolution with realistically decreasing temperature) depends exponentially on the time variable $t'$ used in the simulations at constant temperature (equation~\ref{eqTTprime}). Therefore, for realistic magnetic fields, the time needed to reach the GS equilibrium state would be many orders of magnitude larger than the time to transit to the weak-coupling regime. 
    
\end{enumerate}

Finally, we note some issues that were not included in the present work and might lead to future extensions:
\begin{enumerate}
\item Consistently with our stellar model, we neglected the direct Urca process, which might happen in the inner core of massive NSs, speeding up the thermal evolution and thus strongly reducing the time at which the strong-coupling regime ends. The magneto-chemical evolution would also be 
accelerated proportionally in the inner core, but not in the outer layers, where only modified Urca reactions operate. As in the case considered here, the thermal evolution for realistic magnetic field strengths will still be much faster (by a factor $\gtrsim 8\pi P/B^2$) than the magneto-chemical evolution, therefore the chemical and GS equilibrium should not be reached.

\item As discussed in the introduction, neutrons and protons are expected to become superfluid and superconducting, respectively, relatively early in the NS life, when the core is at temperatures $T\sim 10^{8}-10^{10}\,\text{K}$ \citep{migdal1959superfluidity,page2013stellar}. The inclusion of these effects to study the long-term magneto-thermal evolution can be highly complex as the macroscopic manifestation of the interactions between quantized neutron vortices and magnetic flux tubes needs to be taken into account. Qualitatively, one can anticipate that the approach to chemical equilibrium would be slowed down as Cooper pairing drastically reduces the reaction rates \citep{Yakovlev2001PhR, yakovlev2004neutron}. Additionally, it impacts the cooling rate, initially accelerating it through Cooper pair breaking and formation processes, and eventually delaying it through the suppression of the Urca reactions. Furthermore, disregarding the interactions between vortices and flux tubes, superfluidity and superconductivity suppress inter-particle collisions, potentially resulting in faster ambipolar diffusion. This, in turn, may cause the transition from the strong-coupling regime to the weak-coupling regime to occur earlier. Therefore, including these effects might reinforce the conclusion that no substantial magnetic field evolution will occur during the strong-coupling regime.
\item Throughout this work, for simplicity we have used Newtonian gravity. However, 
the inclusion of general relativity can have significant implications for the overall magneto-thermal evolution. Specifically, when accounting for it in the thermal evolution, it becomes necessary to distinguish between local values of the temperature and the luminosities and their redshifted values as detected by a distant observer. In particular, the two temperatures differ by a factor $e^{-\psi} \sim 1.3$, so the equilibrium neutrino luminosity due to Urca reactions, which is proportional to $T^8$ could change by a non-negligible factor $e^{-8\psi} \approx 8.2$ depending on which temperature is used, with an important quantitative effect on the thermal history (see e.~g., Fig.~7 and Fig.~8 in \citealt{Ofengeim2017NeutrinoLum}). 
\end{enumerate}


\section*{Acknowledgements}
The authors are grateful to D. D. Ofengeim for useful discussions. N. A. M. thanks for the support of ANID doctoral fellowship 21210909 and is grateful to the Department of Applied Mathematics at the University of Leeds. We also thank for the support of FONDECYT Projects 11230837 (F.C.), 1201582 (A.R.), and 1190703 (J.A.V.), and the Center for Astrophysics and Associated Technologies (CATA; ANID Basal project FB210003). J.A.V. also thanks for the support of the Center for the Development of Nanoscience and Nanotechnology (CEDENNA) under CONICYT/ANID grant FB0807. M.E.G. acknowledges support from the Russian Science Foundation [Grant No. 22-12-00048] and is grateful to the Department of Particle Physics \& Astrophysics at the Weizmann Institute of Science for hospitality.

\section*{Data Availability}
The data underlying this article will be shared on reasonable request to the corresponding author.


\bibliographystyle{mnras}
\bibliography{Lib.bib} 



\bsp	
\label{lastpage}

\end{document}